\documentclass[useAMS,usenatbib,usegraphicx]{mn2e}

\title[The Illustris Black Hole Scaling Relations]{The Illustris Simulation: Supermassive Black Hole $-$ Galaxy Connection Beyond the Bulge}

\author[B. Mutlu-Pakdil et al.]{Bur\c{c}in Mutlu-Pakdil$^{1,2,3}$\thanks{E-mail: bmutlupakdil@as.arizona.edu}, Marc S. Seigar$^{1}$, Ian B. Hewitt$^{4}$, Patrick Treuthardt$^{4}$, 
\newauthor
Joel C. Berrier$^{5}$, and Lauren E. Koval$^{6}$\\
$^{1}$Department of Physics \& Astronomy, University of Minnesota Duluth, 1023 University Drive, Duluth, MN 55812-3009, USA\\ 
$^{2}$Minnesota Institute for Astrophysics, University of Minnesota Twin Cities, 106 Pleasant St. SE, Minneapolis, MN 55455, USA\\
$^{3}$Department of Astronomy \& Steward Observatory, 933 North Cherry Avenue, Rm. N204, Tucson, AZ 85721-0065, USA\\
$^{4}$Astronomy \& Astrophysics Research Laboratory, North Carolina Museum of Natural Sciences, 11 W.\ Jones Street, Raleigh,\\
NC 27601, USA\\
$^{5}$Department of Physics \& Astronomy, The University of Nebraska at Kearney, 2504 9th Ave., Kearney, NE 68849, USA\\
$^{6}$Department of Physics \& Astronomy, Oberlin College, 110 N. Professor Street, Oberlin, OH 477074, USA}

\begin{document}
\date{}
\pagerange{\pageref{firstpage}--\pageref{lastpage}} \pubyear{2014}

\maketitle

\label{firstpage}

\begin{abstract}
We study the spiral arm morphology of a sample of the local spiral galaxies in the Illustris simulation and explore the supermassive black hole$-$galaxy 
connection beyond the bulge (e.g., spiral arm pitch angle, total stellar mass, dark matter mass, and total halo mass), finding good agreement with other theoretical 
studies and observational constraints. It is important to study the properties of supermassive black holes and their host galaxies through both observations and 
simulations and compare their results in order to understand their physics and formative histories. 
We find that Illustris prediction for supermassive black hole mass relative to pitch angle is in rather good agreement with observations and that barred and non-barred galaxies follow 
similar scaling relations. Our work shows that Illustris presents very tight correlations between supermassive black hole mass and large-scale properties of the host galaxy, not only for early-type galaxies but also low-mass, blue and star-forming galaxies. These tight relations beyond the bulge suggest that halo properties determine those of a disc galaxy and its supermassive black hole.
 
\end{abstract}

\begin{keywords}
methods: numerical -- cosmology: theory -- cosmology: galaxy formation -- galaxies: spiral
\end{keywords}

\section{Introduction}

It is now well accepted that supermassive black holes (BH) reside at the center of most massive galaxies, both quiescent and active \citep{Kormendy1995,Barth2004,Kormendy2004,Magorrian1998}. 
In the last decades, several scaling relationships between the mass of the central BH and the overall morphology and dynamics of the host galaxy have been found and 
pointed to a co-evolution scenario of galaxy formation and BH growth. Some of these properties, known to correlate well with the BH mass are spheroid velocity dispersion
\citep[$\sigma_{sph}$, e.g.,][]{Gebhardt2000,Ferrarese2000,Tremaine2002,Gultekin2009,McConnell2013}, spheroid luminosity \citep[$L_{sph}$, e.g.,][]{Kormendy1995,Graham2007,Graham2013}, 
spheroid stellar mass \citep[$M_{sph}$, e.g.,][]{Kormendy1995,Magorrian1998,Haring2004,Savorgnan2016}, the central stellar concentration of the spheroid \citep{Graham2001}, 
S\'{e}rsic index ($n$) of the major-axis surface brightness profile \citep[e.g.,][]{Graham2003,GrahamDriver2007,Savorgnan2013}, and spiral arm pitch angle 
\citep[$P$: a measure of the tightness of spiral arms,][]{Seigar2008,Berrier2013}. A common interpretation for these relationships is that BHs regulate their own growth 
and that of their host galaxies through AGN-feedback \citep{SilkRees1998,Springel2005,Bower2006,Croton2006,DiMatteo2008,Ciotti2009,Fanidakis2011}. Therefore, these scaling 
relations have important implications not only for BHs and galaxies but also for understanding the effects of AGN feedback.

Recent observational studies have significantly revised these scaling relations, and reported that galaxies with different properties (e.g., pseudo-bulges versus real bulges, barred 
versus non-barred, or cored versus power-law ellipticals) may correlate differently with their central BH masses \citep[e.g.,][]{Graham2008,McConnell2013,KormendyHo2013,Gebhardt2011}.
Therefore, it is of fundamental importance to investigate the supermassive black hole$-$galaxy connection from a theoretical point of view and also investigate if supermassive 
black hole$-$galaxy co-evolution occurs for all galaxy types or if different galaxy types hold weaker or stronger physical links with their BHs. 
Although several cosmological simulations have investigated the supermassive black hole$-$galaxy connection and confirmed the establishment of the scaling relations 
\citep[e.g.,][]{Sijacki2007,DiMatteo2008,BoothSchaye2009,Dubois2012,Hirschmann2014,Khandai2015,Schaye2015}, 
the co-evolution of the BHs and their galaxies in different galaxy types has not been studied in detail. This is due to the difficulty in simulating representative galaxy samples 
with significant spatial resolution. Recently, the Illustris simulation project \citep[e.g.,][] {Vogelsberger2014,Genel2014} has successfully simulated representative galaxy samples covering the 
observed range of morphologies and provided good spatial resolution to resolve the basic structural properties of the host galaxies. This has made it possible to study the supermassive black 
hole$-$galaxy connection in detail. 

Illustris is a large-scale and high resolution hydrodynamic simulation, with a broad range of astrophysical processes: gas cooling with
radiative self-shielding corrections, energetic feedback from growing BHs and exploding supernovae, stellar evolution and associated chemical enrichment and stellar mass 
loss, and radiation proximity effects for AGN \citep[see][for details]{Vogelsberger2014}. By using a small subset of 42 spiral galaxies, \citet{Vogelsberger2014} demonstrated that Illustris
reproduces the observed stellar and baryonic Tully-Fisher relation \citep{TullyFisher1977,McGaugh2012} reasonably well. \citet{Sijacki2015} presented results on the BH scaling relations 
relative to bulge ($M_{sph}$ and $\sigma_{sph}$) from the Illustris simulation. They concluded that BHs and galaxies co-evolve at the massive end, but there is no tight relation with 
their central BH masses for low-mass, blue and star-forming galaxies. Here, we study the BH scaling relations beyond the bulge, i.e., spiral arm pitch angle, total stellar mass ($M_{\star,total}$),
dark matter mass ($M_{DM}$), and total halo mass ($M_{halo}$) to complement the study of \citet{Sijacki2015}. We compare the Illustris predictions with other theoretical results and 
observational constraints to further understanding of the supermassive black hole $-$ galaxy connection. To do so, we first concentrate on a randomly selected sample of the 
Illustris spiral galaxies that do not have rings, and study the face-on spiral arm morphology in multiple wavebands ($B$, $R$ and $K$). We explore its connection with BH and DM for 
our sample of spiral galaxies in Illustris. Then, we focus on the redshift $z=0$ central galaxies with $10.0<\log(M_{\star,2R}/M_{\sun})<13.0$, where $M_{\star,2R}$ is the stellar mass within 
the 2x stellar half-mass radius, and study the Illustris predictions relative to the $M_{*,total}$, $M_{DM}$ and $M_{halo}$ of the host galaxy. Note that \citet{Snyder2015} presented 
the Illustris predictions of the $M_{BH}-M_{*,total}$ and $M_{BH}-M_{DM}$ relationships in the context of their implications for galaxy morphology. They investigated how Illustris galaxy morphologies 
depend on other aspects of the simulated galaxies, such as their optical sizes, BH masses, and DM halo masses. However, our aim is to first clearly understand these 
different relations (e.g., which one may be the strongest correlation), especially for low-mass, blue and star-forming galaxies. 

This paper is organized as follows. In Section 2 we concentrate on a sample of the simulated spiral galaxies to study the spiral arm morphology in multi-wavebands ($\S$2.1) 
and explore its connection with the BH and DM masses ($\S$2.2). In Section 3, we study the Illustris predictions relative to the $M_{*,total}$ ($\S$3.1), $M_{DM}$ ($\S$3.2) and $M_{halo}$ ($\S$3.3) of 
the host galaxies by comparing with other theoretical results and observational constraints. We finally discuss our results and draw conclusions in Section 4.

\section{The Illustris Spiral Arm Morphology and Pitch Angle Scaling Relations}
\subsection{Spiral Arm Morphology in Multi-wavebands}
The Illustris Project consists of hydrodynamical simulations of galaxy formation in a box of $75$ $Mpc$ $h^{-1}$ ($= 106.5$ $Mpc$) on the side \citep{Vogelsberger2014,Genel2014}. 
This volume was simulated with and without baryons at several resolution levels, here we present results from the simulation with the highest resolution (with $N_{DM}=1820^3$ DM particles 
and $N_{baryon}=1820^3$ baryon resolution elements) which resolves baryonic matter with mass $1.26\times10^6 M_{\sun}$. For all calculations, we adopt the cosmology used 
for the Illustris simulation, that is cosmological parameters consistent with the latest Wilkinson Microwave Anisotropy Probe (WMAP)-9 measurements ($\Omega_{M}=0.2726$, $\Omega_{\lambda}=0.7274$,
$\Omega_{b}=0.0456$, $\sigma_{8}=0.809$, and $H_{0}=70.4$ $km$ $s^{-1}$ $Mpc^{-1}$ with $h=0.704$). Parameters in Illustris were tuned to match the redshift $z=0$ stellar mass and halo 
occupation functions, and evolving cosmic star formation rate density \citep{Vogelsberger2014}, and \citet{Torrey2015} showed that this parameterization produces galaxy populations 
that evolve consistently with observations.

In this section, we focus on the $z=0$ spiral galaxies in Illustris to study the spiral arm morphology in multiple wavebands and explore its connection with BH and DM. 
\citet{Snyder2015} defined a non-parametric bulge-strength parameter, $F(G,M_{20}$), to serve as a rough automated assessment of morphological types for the $z=0$ galaxies in Illustris. 
In this parameter, $G$, is the relative distribution of the galaxy pixel flux values (the Gini coefficient) and $M_{20}$ is the second-order moment of the brightest 20$\%$ of the galaxy's flux 
\citep{Conselice2003,Lotz2004}. They found that late-type galaxies reside in the ($M_{20}<-2$, $G<0.6$) region at $10.0<\log(M_{\star,2R}/M_{\sun})<11.0$. Following their definition for 
late-type galaxies, we first select 5131 galaxies from their $g-$band morphology catalog, based on their location in the $G-M_{20}$ plane. This sample includes a large number of 
galaxies with a ``ring-like'' morphology or an irregular/disturbed shape \citep{Vogelsberger2014,Snyder2015}. About 10 percent of disc galaxies in the mass range 
$M_{\star,2R}\sim10^{10.5-11} M_{\sun}$ exhibit strong stellar and gaseous ring-like features which are likely not realistic \citep{Snyder2015}. 
Hence, we use the face-on synthetic images in the Johnson-$B$ (444.9 nm) filter to construct a subsample of 2258 blue galaxies that do not have rings but display convincing evidence 
of definable spiral structure from image inspection. The methods of measuring spiral arm pitch angle are currently very much dependent on the visual inspection conducted by the user \citep{Davis2012} 
and that makes a challenge to measure pitch angles for a large sample of the simulated galaxies. Therefore, we randomly select 95 galaxies out of our subsample and only attempt pitch 
angle measurements on these randomly selected galaxies. After constructing our spiral sample, we study the spiral arm morphology by using the face-on synthetic images in the Johnson-$B$ (444.9 nm), Cousins-$R$ (659.9 nm) and Johnson-$K$ (2.2 $\micron$) filters, which are available at the Illustris webpage (http://www.illustris-project.org/galaxy\_obs/) \citep{Torrey2015Image}. For brevity, 
we only use the data reported for the face-on viewing direction, i.e. camera index $= 3$, since our algorithm requires image deprojection as this produces the most accurate results \citep{Davis2012}. Figure 1 shows example images in the $B$, $R$, and $K$ filters (from left to right) of the $z=0$ Illustris galaxies, arranged by increasing stellar mass from top ($M_{\star,2R} \sim 10^{10} M_{\sun}$) 
to bottom ($M_{\star,2R} \sim 10^{11} M_{\sun}$). The remainder of this section analyzes measurements of synthetic images that are most like those in this figure.

\begin{figure*}
  \includegraphics[width=5.5cm]{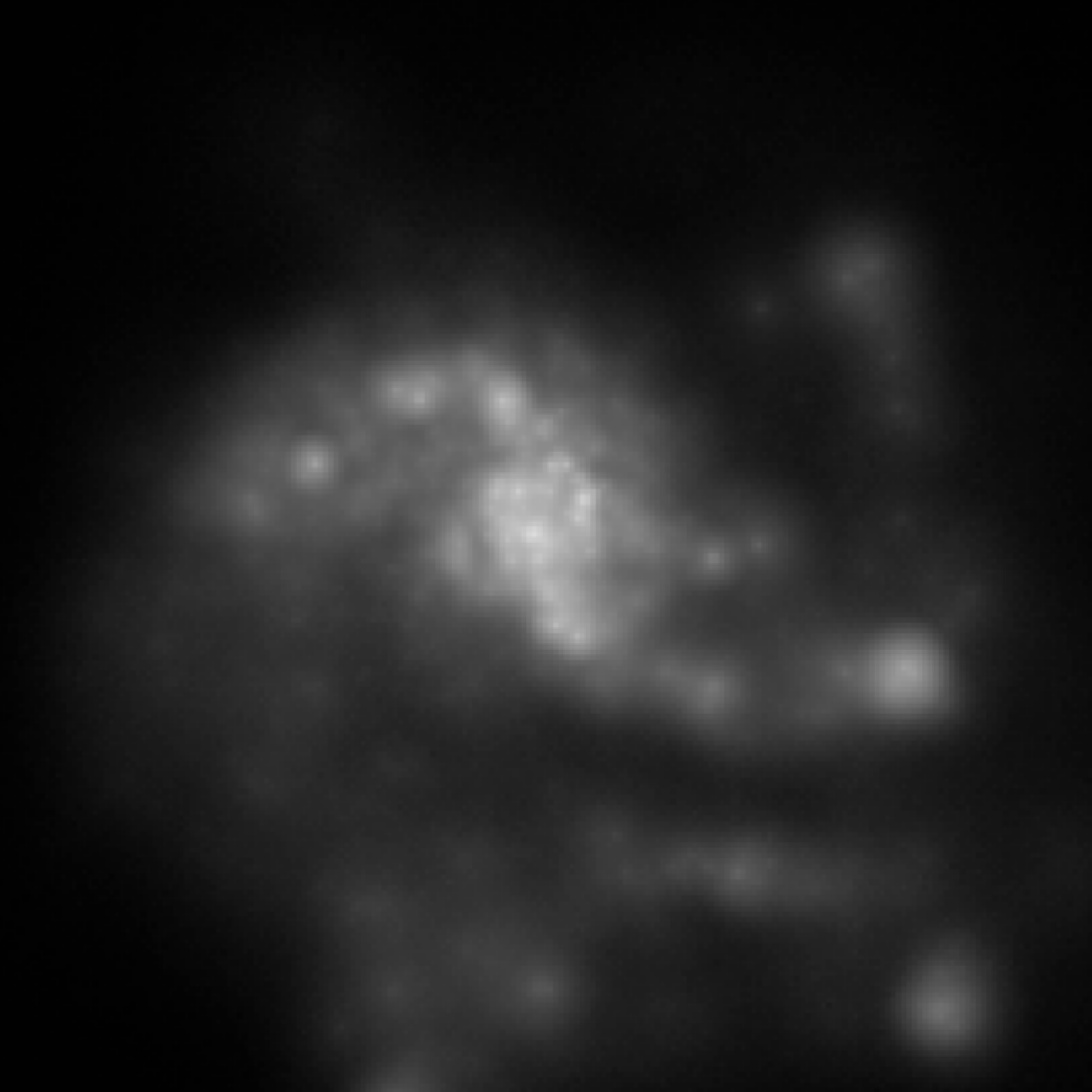}
  \includegraphics[width=5.5cm]{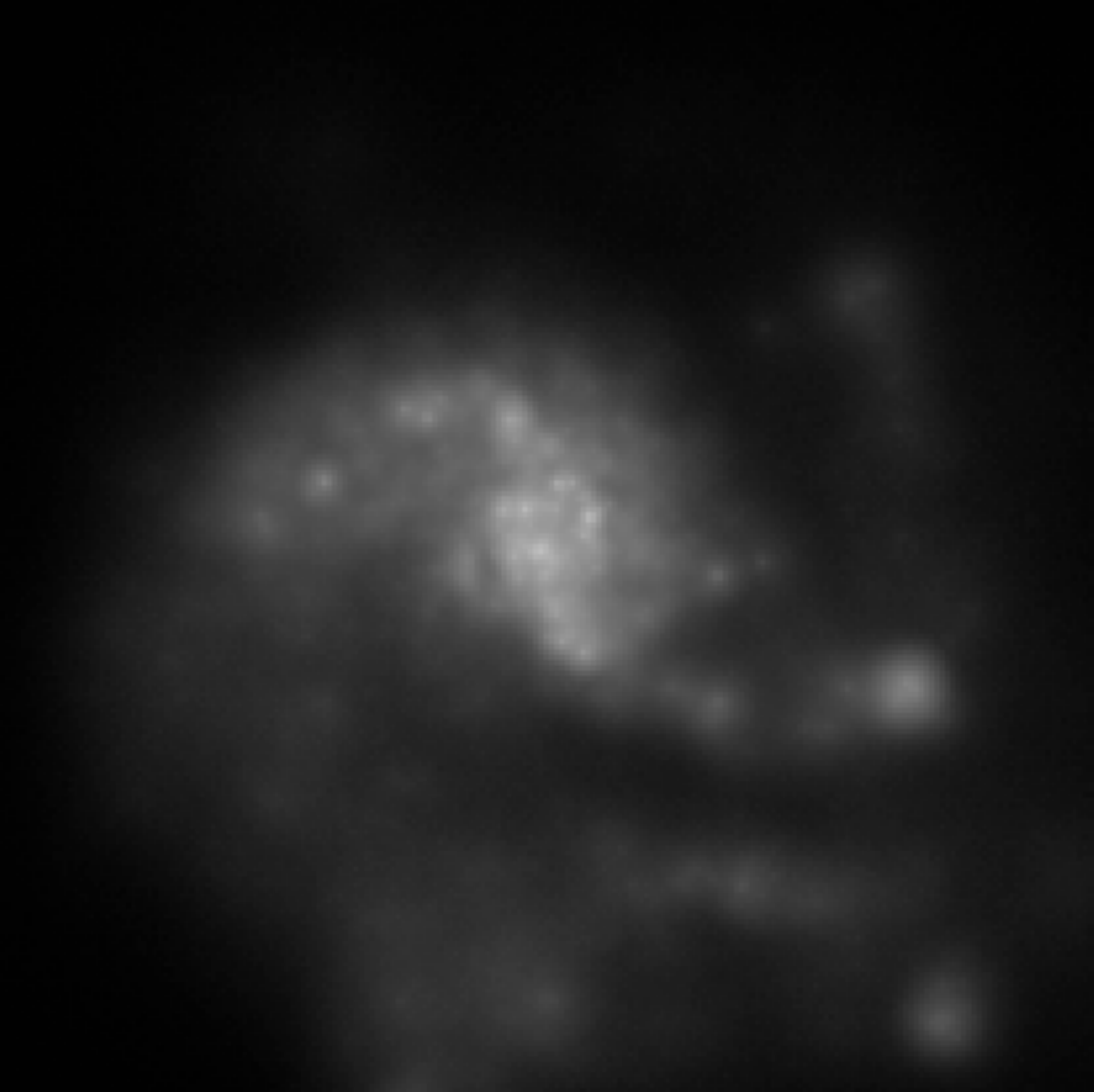}
  \includegraphics[width=5.5cm]{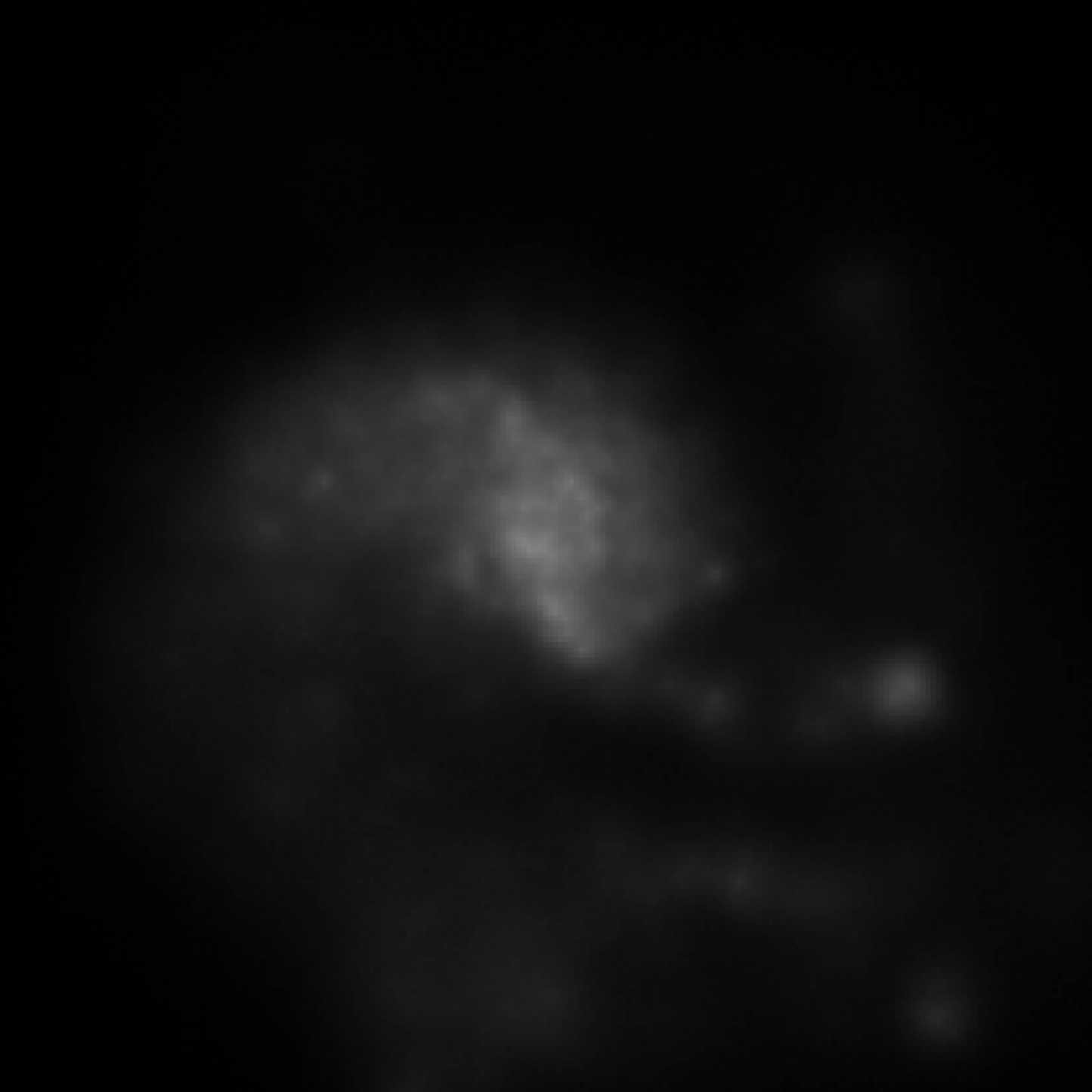}
  \includegraphics[width=5.5cm]{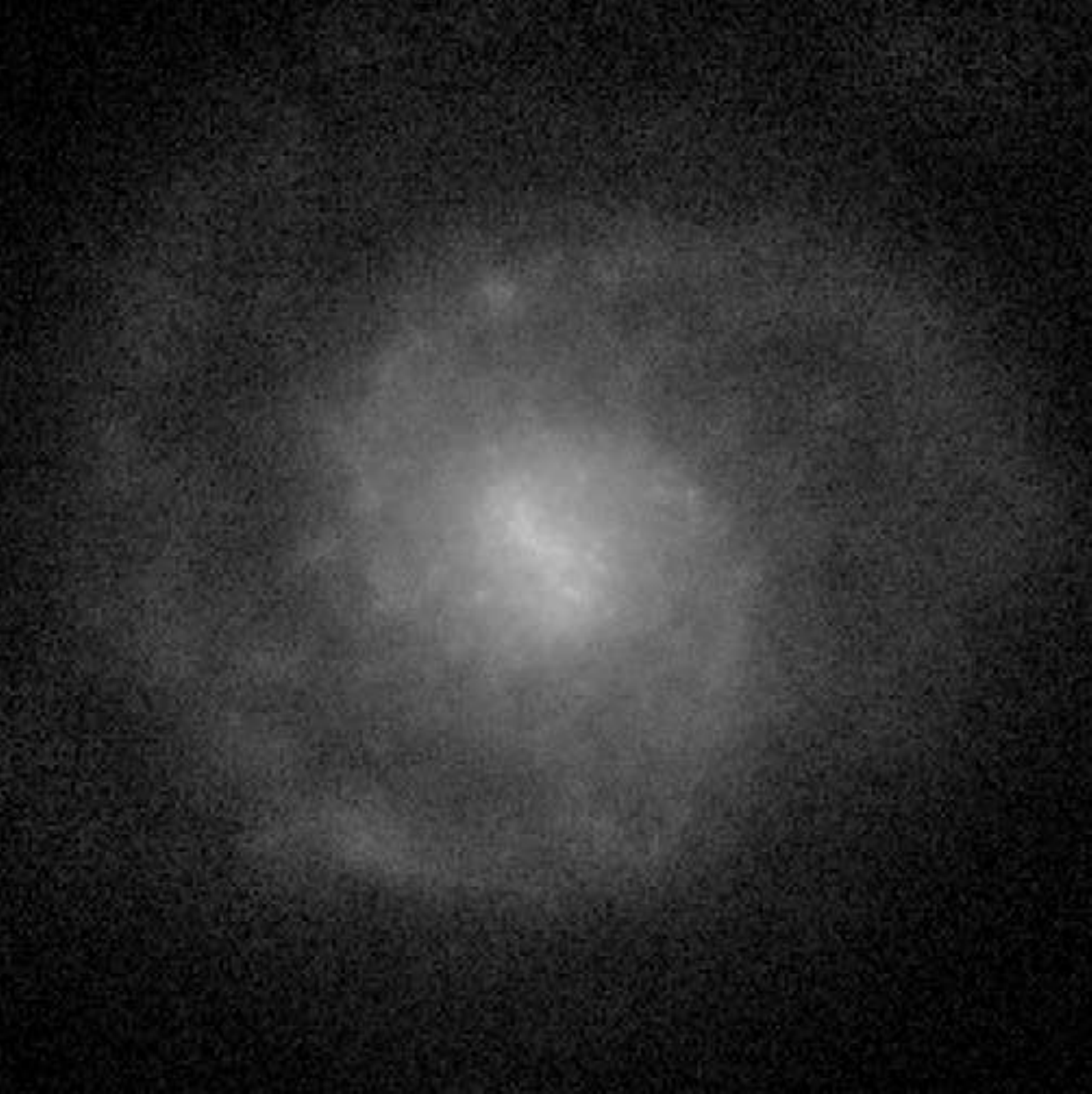}
  \includegraphics[width=5.5cm]{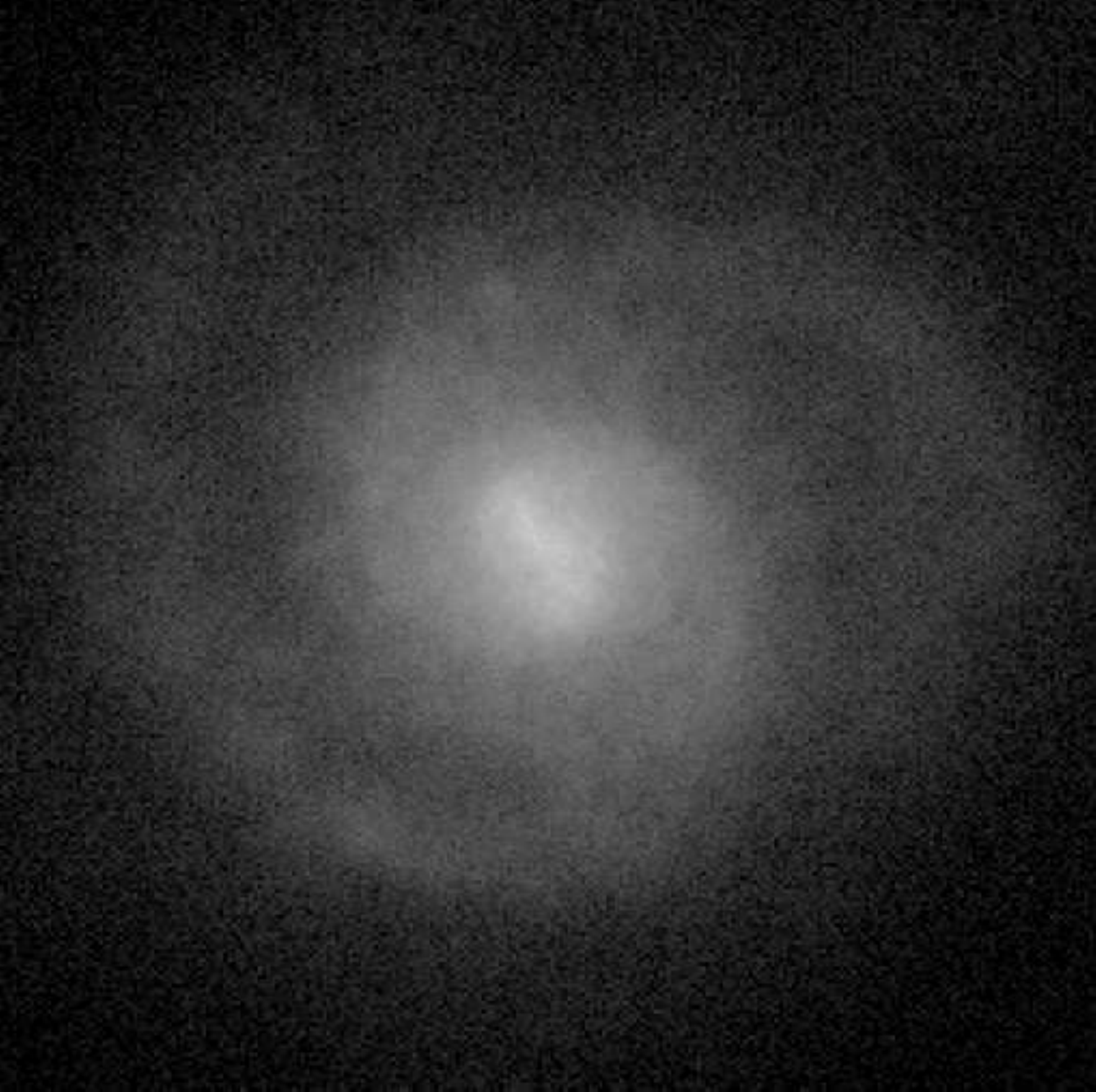}
  \includegraphics[width=5.5cm]{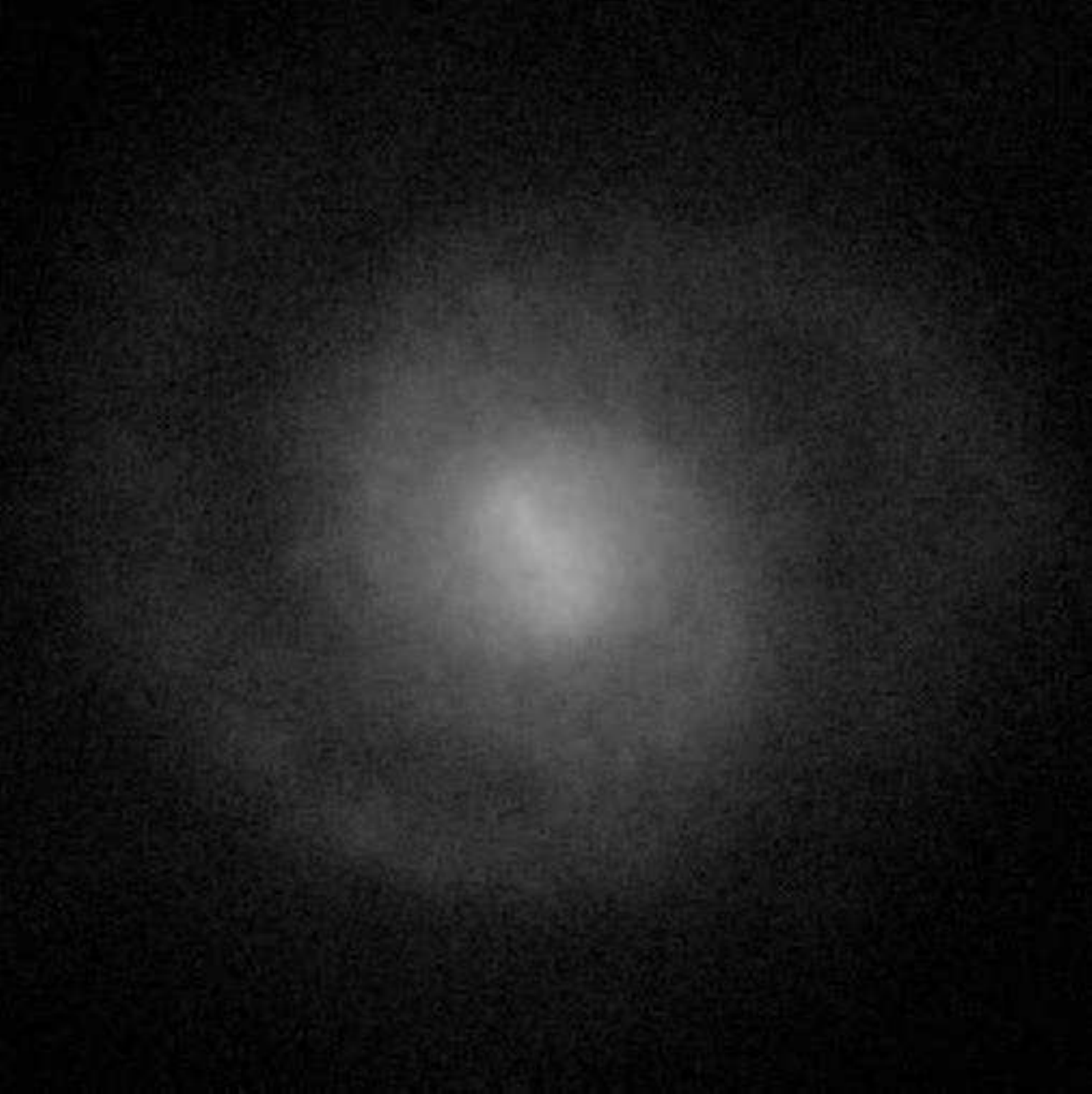}
  \includegraphics[width=5.5cm]{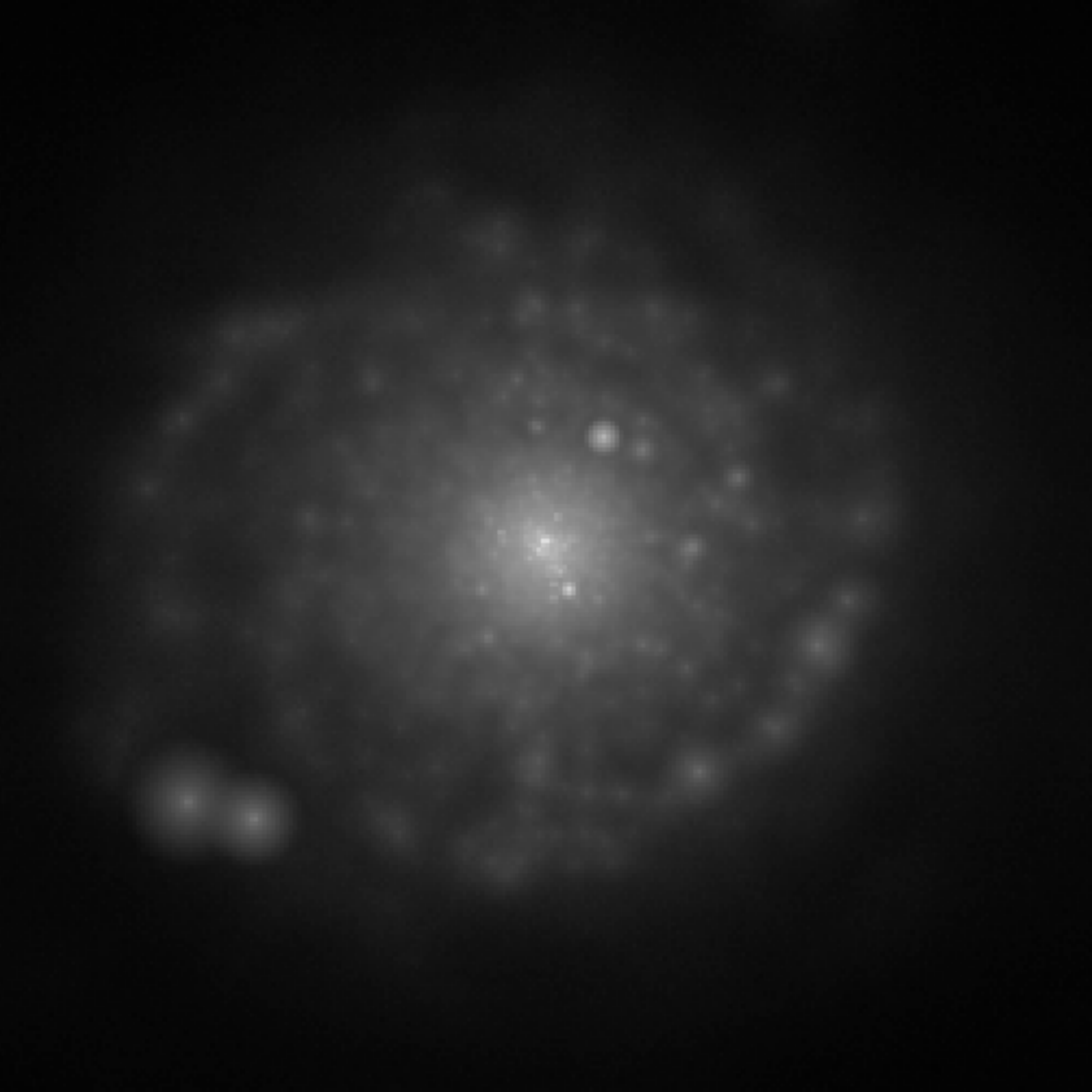}
  \includegraphics[width=5.5cm]{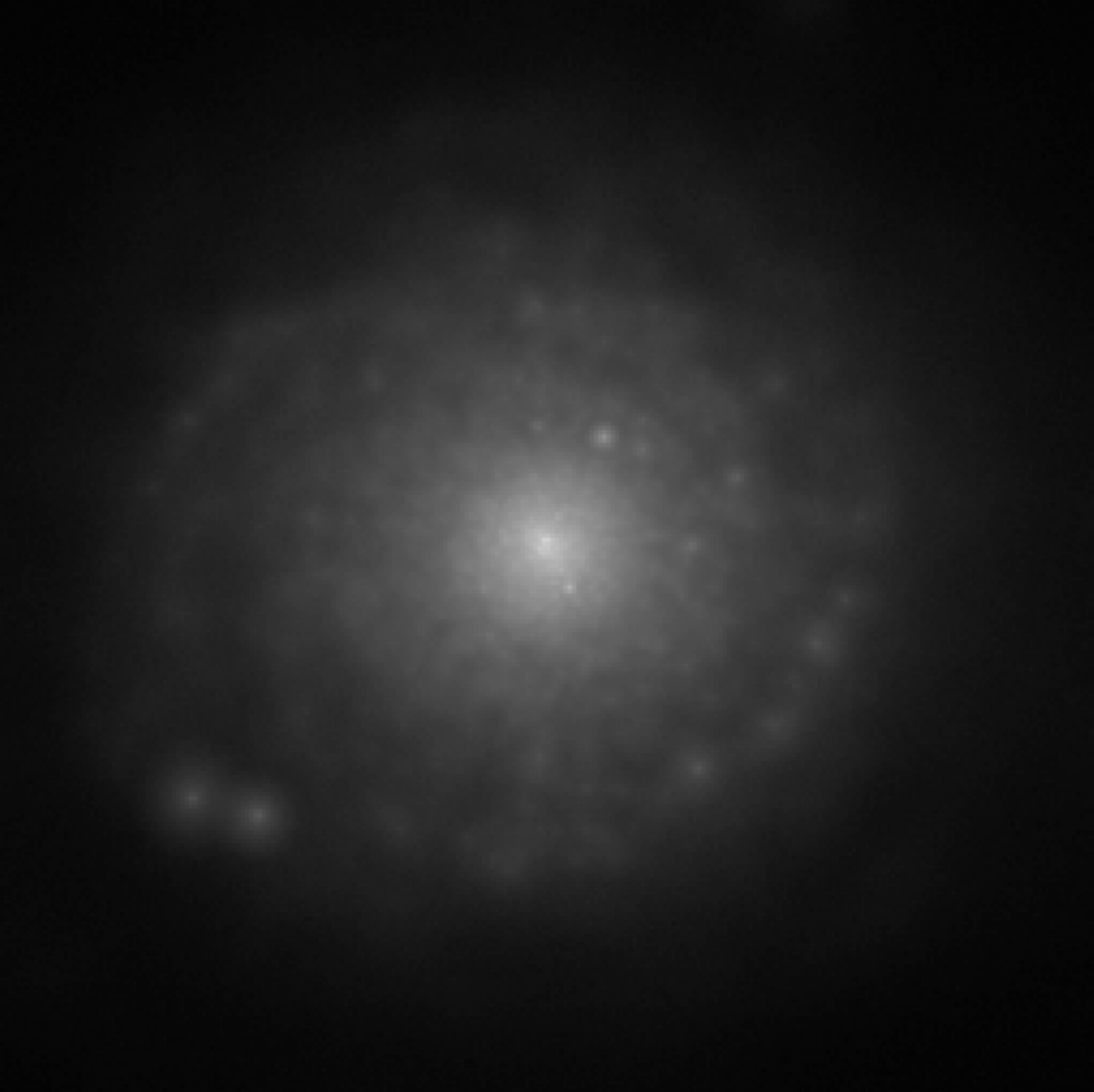}
  \includegraphics[width=5.5cm]{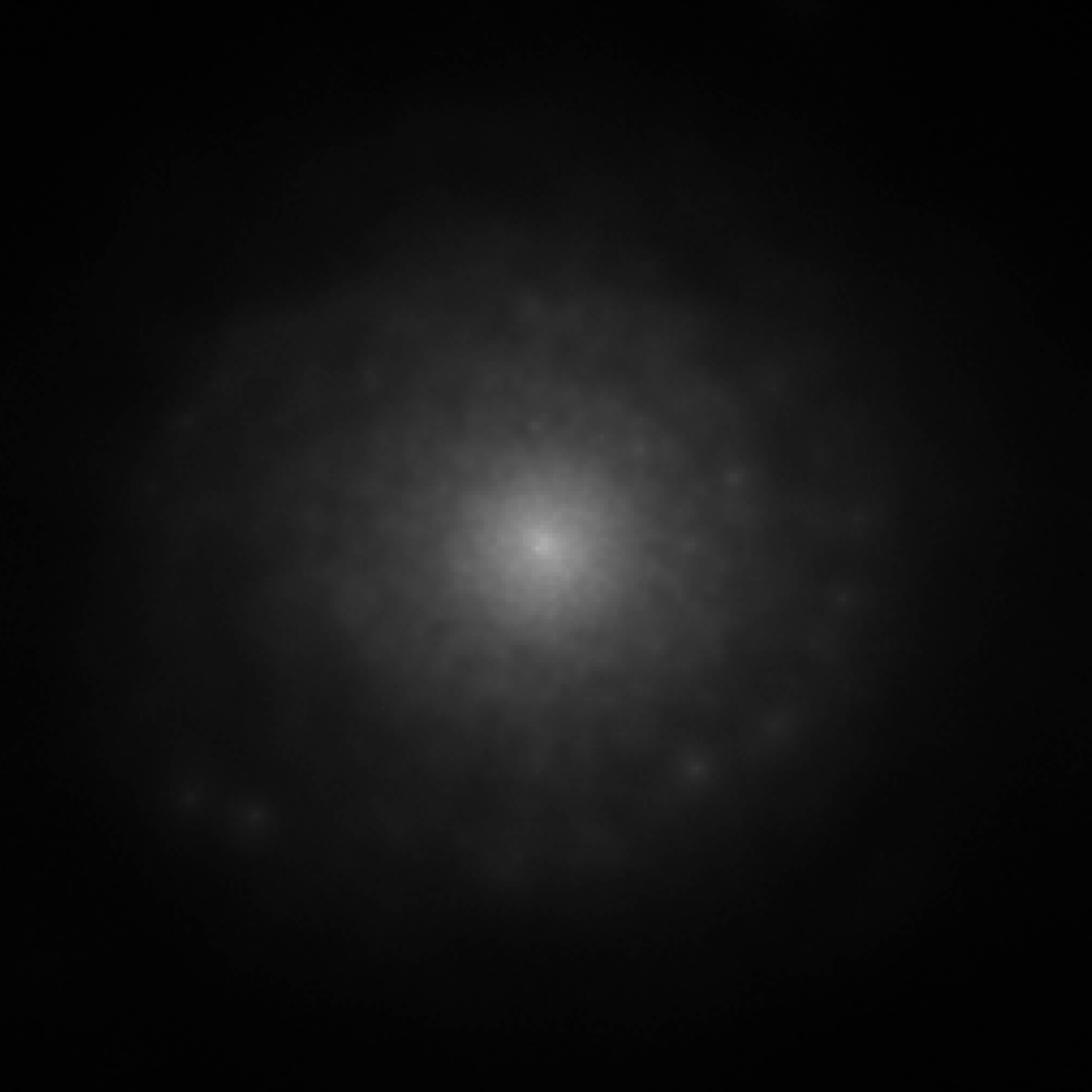}
  \caption{Example images in the $B$, $R$, and $K$ filters of the $z=0$ Illustris galaxies are shown from left to right, respectively. Top: $ID=450471$, Middle: $ID=41098$, 
Bottom: $ID=339972$. They are arranged by increasing stellar mass from top ($M_{\star,2R} \sim 10^{10} M_{\sun}$) to bottom ($M_{\star,2R} \sim 10^{11} M_{\sun}$). 
Note that each image has a size of 256 pixels.}
  \label{FIGURE-1}
\end{figure*}

For our measurements, we employ the two-dimensional (2D) fast Fourier transform (FFT) software called P2DFFT\footnote{P2DFFT is available for download at https://treuthardt.github.io/P2DFFT/}. 
This software is an updated version of 2DFFT \citep{Davis2012} that implements parallel processing, allowing it to run faster than the serial 2DFFT, and has simplified the user input. 
See \citet{Schroeder1994,Puerari1992,Puerari2000} for more details on the 2DFFT algorithm. P2DFFT has been written to allow direct input of FITS images and optionally output inverse Fourier 
transform FITS images. Also included in the package is the ability to generate idealized logarithmic spiral test images of a specified size that have 1 to 6 arms with pitch angles of -75\degr 
to 75\degr. P2DFFT further includes a Python code that outputs Fourier amplitude versus inner radius and pitch angle versus inner radius for each Fourier component ($0 \le m \le 6$). This Python 
routine also calculates the Fourier amplitude weighted mean pitch angle across $1 \le m \le 6$ versus inner radius. 

Operationally, P2DFFT decomposes galaxy images into logarithmic spirals and determines the pitch angle that maximizes the Fourier amplitudes for each harmonic mode ($m$). This code provides 
a systematic way of excluding barred nuclei from the pitch angle measurement annulus by allowing inner radius to vary, and outputs the pitch angle as a function of inner radius.
By using the face-on synthetic images, we avoid the deprojection process and any assumption related with it. Then, we seek a harmonic mode in which pitch angle stays approximately constant 
over the largest range of radii, and measure pitch angle by averaging over this stable region. The error is determined by considering the size of the stable radius segment relative to the 
galaxy radius as well as the degree to which the stable segment is logarithmic \citep[see][for more detail]{Davis2012}. An example of our pitch angle measurements is shown in Figure 2: 
a stable mean pitch angle is determined for the $m=2$ harmonic mode from a minimum inner radius of 50 pixels to a maximum inner radius of 80 pixels, with an outer radius of 110 pixels. 
Note that the typical image size is 256 pixels. The middle panel shows the amplitude of each Fourier component with a measurement annulus defined by an inner radius of 50 pixels and an 
outer radius of 110 pixels. This clearly displays the dominance of the two-armed spiral harmonic mode. The same procedure is used for all 95 galaxies in the $B$, $R$ and $K$ filters. 
Figure 3 shows the comparison between the histograms of the $A(p_{max},m)$ values, i.e, the highest amplitude of the chosen harmonic mode \citep[see][]{Davis2012}, for our Illustris sample and the observations reported by \citet{Berrier2013}. We perform a Kolmogorov-Smirnov test (K-S test) by using MATLAB \texttt{kstest2} task, which rejects the null hypothesis at the 5\% significance level. This agreement suggests that our comparison is meaningful. We also visually 
classify each galaxy as barred/non-barred according to the presence/absence of a clear sign of a bar structure. Our classification is based on visual inspection of the $K$-band images since 
$NIR$ bands show bars more frequently \citep{Eskridge2000}. We find a majority of our sample consisting of non-barred morphologies: 21 barred and 74 non-barred. This is in good agreement with the percentage of barred galaxies ($29.4\pm0.5$) reported by \citet{Masters2011}.
The pitch angle measurements along with the data related with this paper are presented in Table 1.

\begin{figure*}
\centering
  \includegraphics[width=6.4cm]{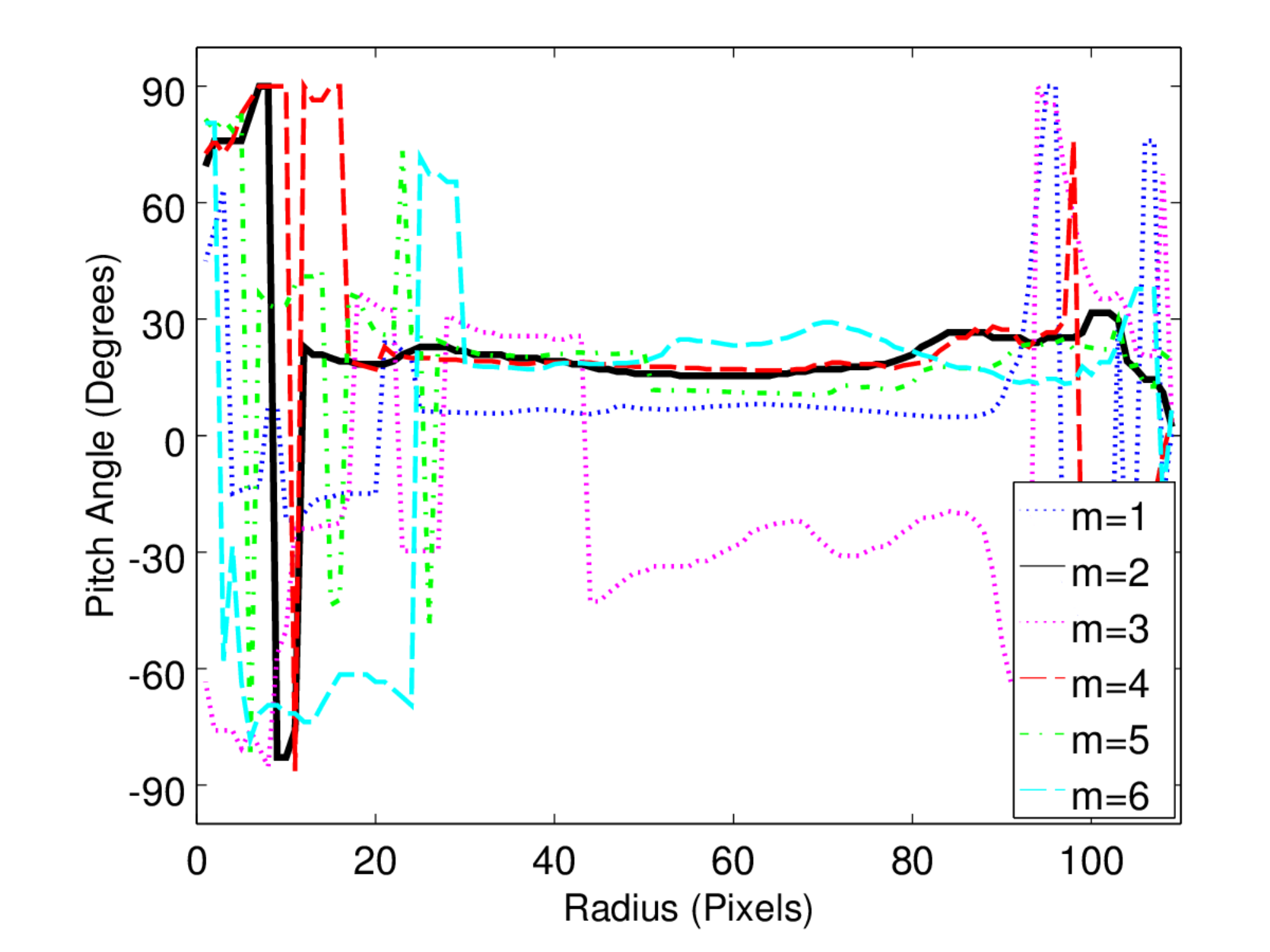}
  \includegraphics[width=6.55cm]{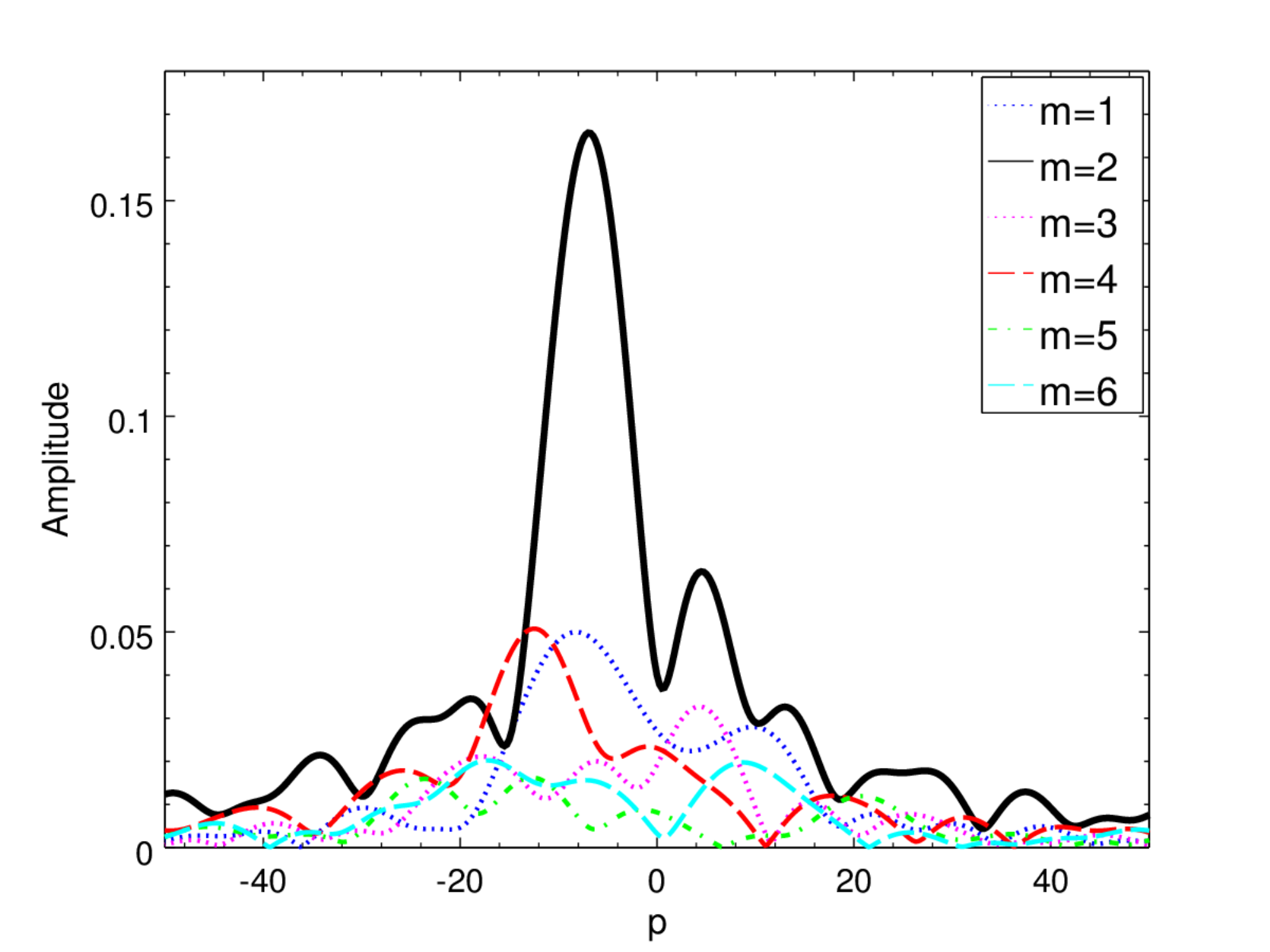}
  \includegraphics[width=4.55cm]{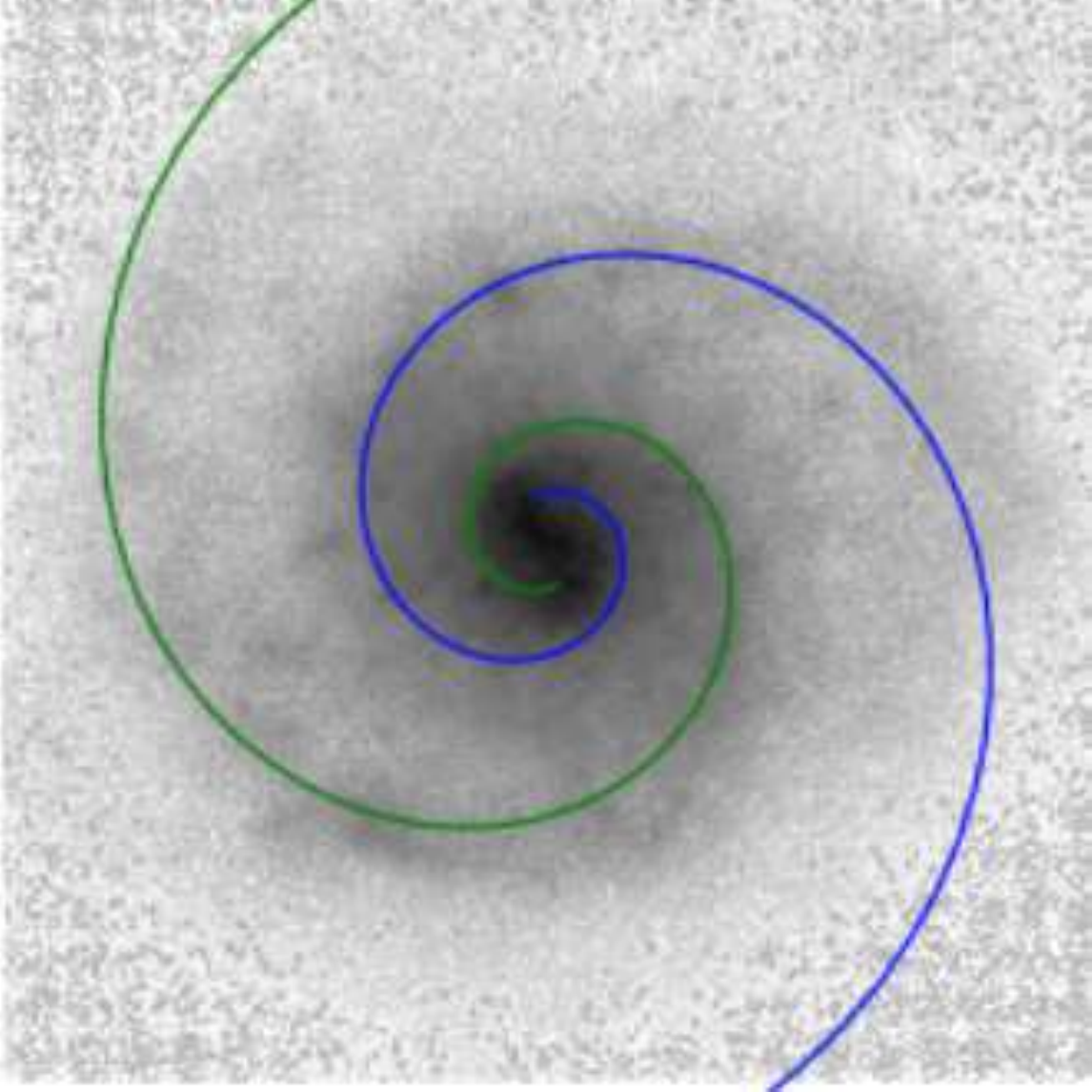}
  \caption{Left: Example $B$-band pitch angle profile of a simulated spiral galaxy ($ID=41098$) from our sample is shown as a function of inner radius. A stable mean pitch angle is determined for
the $m=2$ harmonic mode (solid black line): $P=15.61\pm3.06$ degrees. Middle: The amplitude of each Fourier component for the $B$-band image of this simulated galaxy with a measurement annulus 
defined by an inner radius of 50 pixels and outer radius of 110 pixels. Right: An overlay image of a 15.61 degrees pitch angle spiral is shown on top of this simulated galaxy.
Note that the image size is 256 pixels.}
  \label{FIGURE-2}
\end{figure*}

\begin{figure*}
\centering
  \includegraphics[width=10cm]{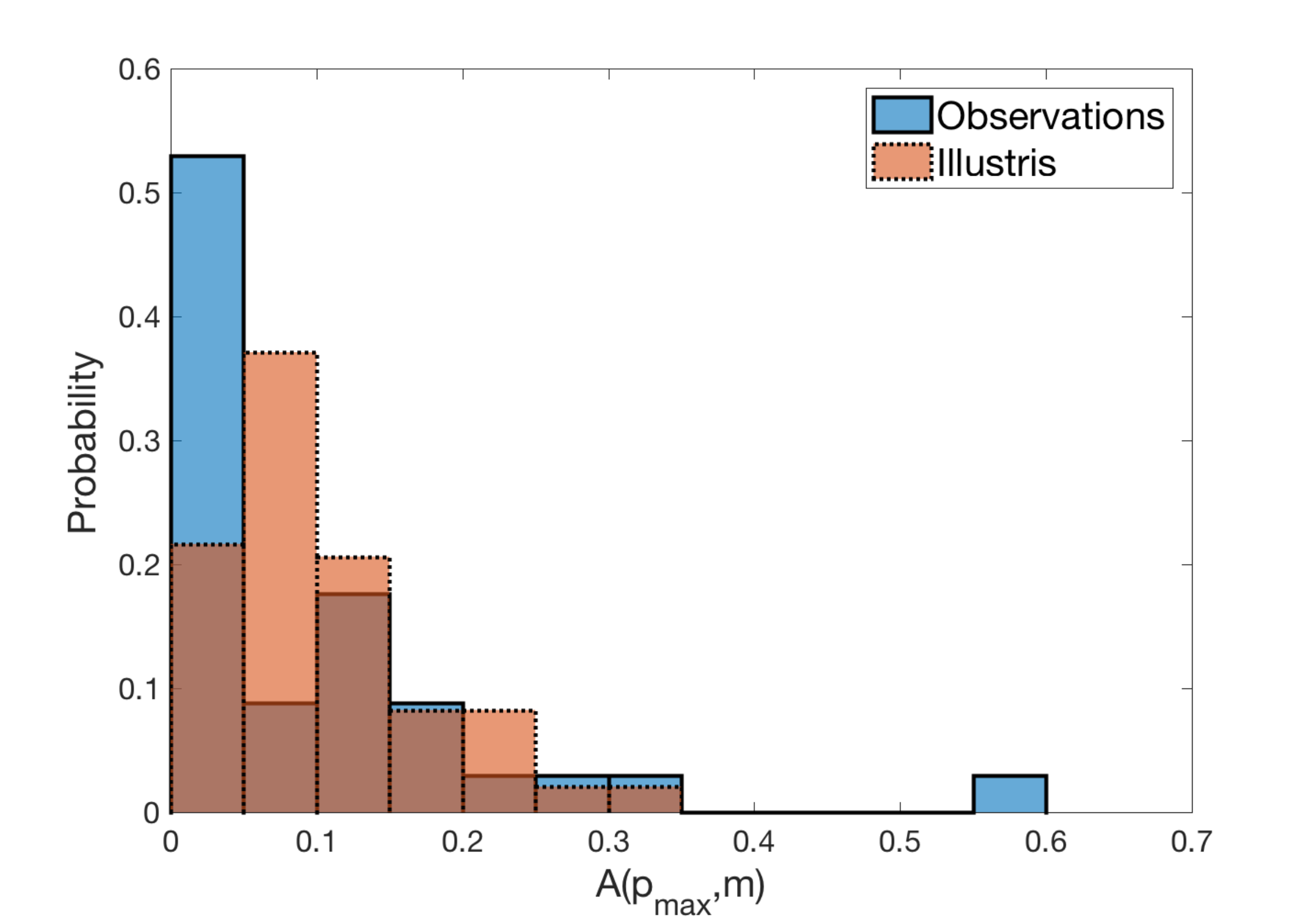}
  \caption{Comparison of the histograms of the $A(p_{max},m)$ values of our Illustris sample and the observed sample from \citet{Berrier2013} is shown. Our K-S test rejects 
the null hypothesis at the 5\% significance level. This agreement suggests that our comparison is meaningful.}
  \label{FIGURE-3}
\end{figure*}

It is important to investigate the possibility of different pitch angles arising in different wavebands of light. While optical $B$-band images tend to trace the bright massive star-forming 
regions of a galaxy, near-infrared ($NIR$) images tend to trace the old stellar populations (and thus the spiral density wave) in galaxies \citep{Seigar1998,Eskridge2002}. 
Moreover, a flocculent spiral structure in the $B$-band may appear as a grand-design in the $NIR$ imaging \citep{Thornley1996}. \citet{Seigar2006} demonstrated the existence of a 1:1 relation 
between the $B$- and $NIR$-band (either $Ks$ or $H$) pitch angles for a sample of 66 galaxies from a combination of the Carnegie-Irvine Galaxy Survey \citep{Ho2011} and the Ohio State
University Bright Spiral Galaxy Survey \citep{Eskridge2002}. Later, \citet{Davis2012} confirmed this relation (between $B$ and $I$) by remeasuring a subset of 47 galaxies of the galaxies 
appearing in \citet{Seigar2006}. Note that \citet{Seigar2006} used an earlier version of the 2DFFT method and \citet{Davis2012} used a method similar to ours. Recently, \citet{PourImani2016} found 
a modest difference between the bands at the extreme ends of this range, from $B$ to 3.6$\micron$, but concluded that their results are consistent with the results 
of \citet{Seigar2006} and \citet{Davis2012} because of the small difference they recovered. Here we carefully investigate the simulated spiral arms when viewed in different 
wavebands and check if Illustris estaiblishes a 1:1 relation that is expected from the observations. Figure 4 presents our pitch angle measurements in different wavebands, showing that pitch angles
are similar whether measured in the optical or $NIR$ regimes. Although there are small-scale differences between spiral arms in different wavebands of the optical-NIR spectrum, the absence 
of a systematic behaviour in pitch angle values (such as being above or below relative to the 1:1 relation line) is consistent with the observational findings of \citet{Seigar2006} and 
\citet{Davis2012}. The overall structure of the simulated spiral arms is consistent across the optical-NIR spectrum. Furthermore, neither barred (green stars) nor non-barred galaxies 
(black circles) show a significant difference in pitch angle in different wavebands. This implies that pitch angle derived is not biased by a presence of a bar. In addition, the most 
common $K$-band pitch angle in our sample is 21$\degr$, which is very close to the most probable pitch angle of 18.52$\degr$ that \citet{Davis2014} derived from a statistically complete 
collection of the brightest spiral galaxies.

\begin{figure*}
  \includegraphics[width=8.8cm]{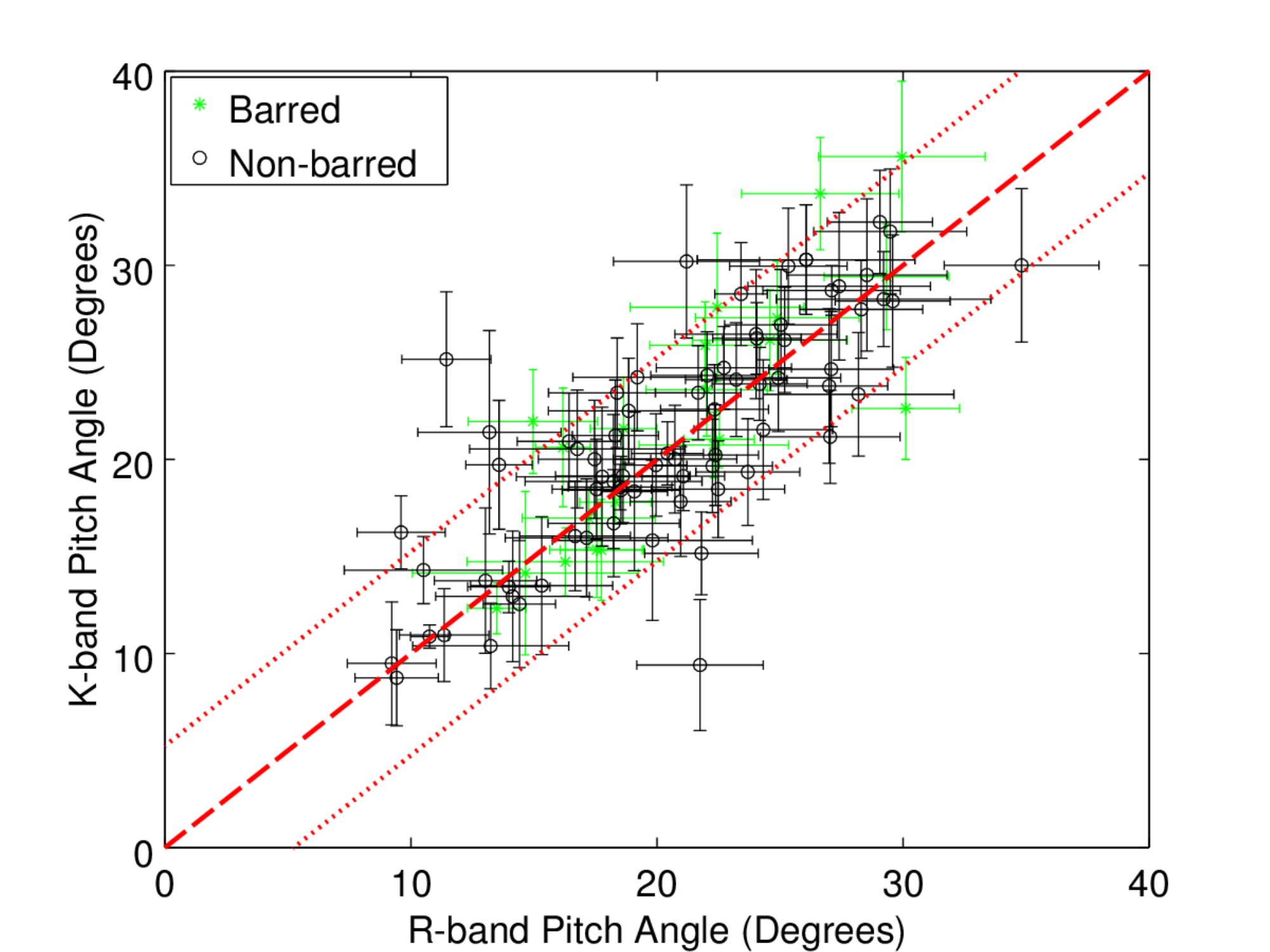}
  \includegraphics[width=8.8cm]{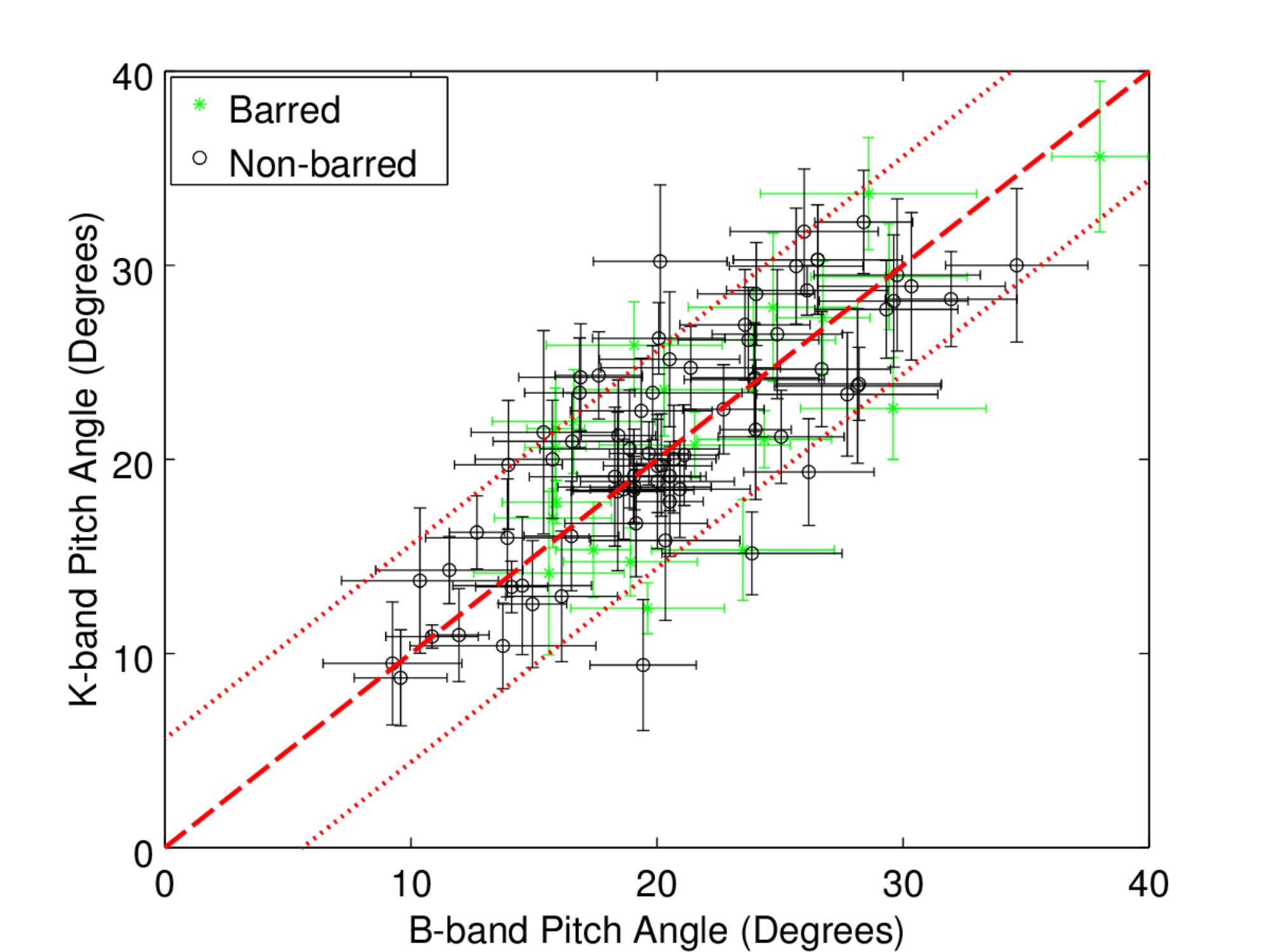}
  \caption{The pitch angle measurements for our spiral sample are shown in different wavebands (Left: $K$ versus $R$, Right: $K$ versus $B$). Galaxies with bars are shown by green stars.
Galaxies with no clear sign of a bar are shown by black circles. The total RMS scatter is plotted in dotted red lines above and below the 1:1 relation line (dashed red line): 5.24\degr in $K$ 
versus $R$ and 5.60\degr in $K$ versus $B$. Although there are small-scale differences between spiral arms in different wavebands of the optical-NIR spectrum, 
the overall structure of the simulated spiral arms is consistent across the optical-NIR spectrum, which is consistent with the observational findings of \citet{Seigar2006} and \citet{Davis2012}.}
  \label{FIGURE-4}
\end{figure*}

\subsection{Pitch Angle Scaling Relations}
Observational evidence shows a tight correlation between pitch angle and the central BH mass in disc galaxies \citep{Seigar2006,Berrier2013}.
It has been shown empirically that there is a link between pitch angle and the central mass concentration of galaxies \citep{Seigar2005,Seigar2006}. Also, 
the supermassive black hole$-$bulge connection is widely established as a result of observed correlations of the BH 
mass with $\sigma_{sph}$, $M_{sph}$ and $L_{sph}$. Since both the BH mass and pitch angle are intimately related to the central mass concentration, it is no surprise that 
they correlate with each other quite strongly. For this relation, \citet{Berrier2013} reported a scatter less than 0.48 dex, which is lower than the intrinsic 
scatter ($\approx0.56$ dex) of the $M_{BH}-\sigma_{sph}$ relation, using only late-type galaxies \citep{Gultekin2009}. While pitch angle depends inversely on central bulge mass 
in bulge-dominated galaxies, it correlates to relative concentration of mass toward the galaxy's center in disc-dominated galaxies, especially the extreme case 
of bulgeless galaxies \citep{Berrier2013}. Unfortunately, the relation between the BH mass and galaxy characteristics in disc-dominated galaxies is not clear empirically
since relatively few BH masses have been directly measured in such galaxies. 

We adopt $M_{BH}$ directly from the simulation outputs as the BH mass contained within the stellar half-mass radius. For a given simulated galaxy, we define $M_{DM}$ 
as the total mass of dark matter and $M_{halo}$ as the total mass of all particles (all types) bound to the subhalo. We use the $K$-band pitch angle values in the scaling relations. 
In Figure 5, we show the Illustris prediction (solid blue line) for the BH mass relative to pitch angle for our spiral sample, and compare it with the observed $M_{BH}-P$ relation 
(solid red line) by \citet{Berrier2013}. The agreement between the Illustris result and the observations is very good: the slope and normalization of the observed $\log(M_{BH}/M_{\sun})-P$ 
relation are $-0.062\pm0.009$ and $8.21\pm0.16$, respectively, whereas the simulation predicts $-0.055\pm0.001$ and $8.40\pm0.01$. The Illustris best fit is obtained using the robust linear 
least-squares fitting method (i.e., bisquare weights) in MATLAB Curve Fitting Toolbox (Goodness of fit: R$^{2}=0.36$, RMSE$=0.44$). We use the same fitting method throughout the paper. 

\begin{figure*}
  \includegraphics[width=13cm]{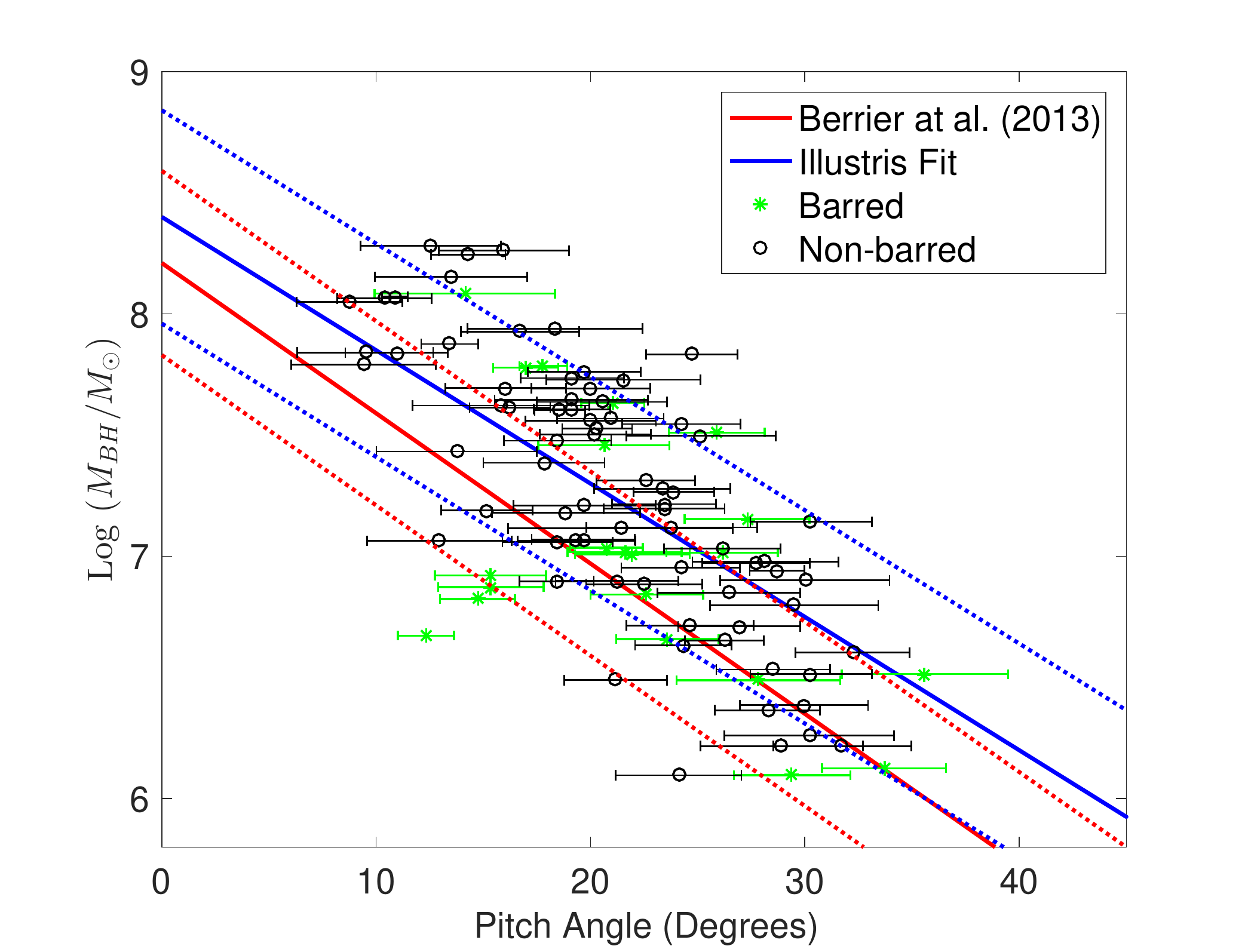} 
  \caption{The Illustris prediction for the BH mass relative to pitch angle for our spiral sample is shown by the solid blue line, which represents the best fit to the data. 
Galaxies with bars are shown by green stars. Galaxies with no clear sign of a bar are shown by black circles. The total RMS scatter of $\pm0.44$ dex in the $\log(M_{BH}/M_{\sun})$ direction 
is plotted in dotted blue lines above and below the best-fit line. For comparison, we have also plotted the observed $M_{BH}-P$ relation by \citet{Berrier2013}, represented by the solid red 
line and bounded by its $\pm0.38$ dex RMS scatter in the $\log(M_{BH}/M_{\sun})$ direction, above and below with dotted red lines. Overall, the Illustris simulation reproduces the observed 
relation very well for disc galaxies.}
  \label{FIGURE-5}
\end{figure*}

In Figure 6, we display the Illustris predictions for DM mass and halo mass relative to pitch angle for our spiral sample. The simulation predicts a very tight correlation between 
these parameters: the slope and normalization are $-0.029\pm0.001$ and $12.46\pm0.01$ (R$^{2}=0.22$, RMSE $=0.30$) for the $\log(M_{DM}/M_{\sun})-P$ relation, respectively, whereas $-0.027\pm0.001$ and 
$12.47\pm0.01$ (R$^{2}=0.21$, RMSE $=0.29$) for the $\log(M_{halo}/M_{\sun})-P$ relation. It is especially interesting that Illustris predicts a tighter correlation between pitch angle and 
DM mass (and halo mass) when compared to the relation between pitch angle and the BH mass. This suggests that pitch angle can be used as a proxy for the DM mass (and halo mass) of disc galaxies.
However, we emphasize that there is a large scatter in the DM mass (and halo mass) at fixed pitch angle.

\begin{figure*}
  \includegraphics[width=\columnwidth]{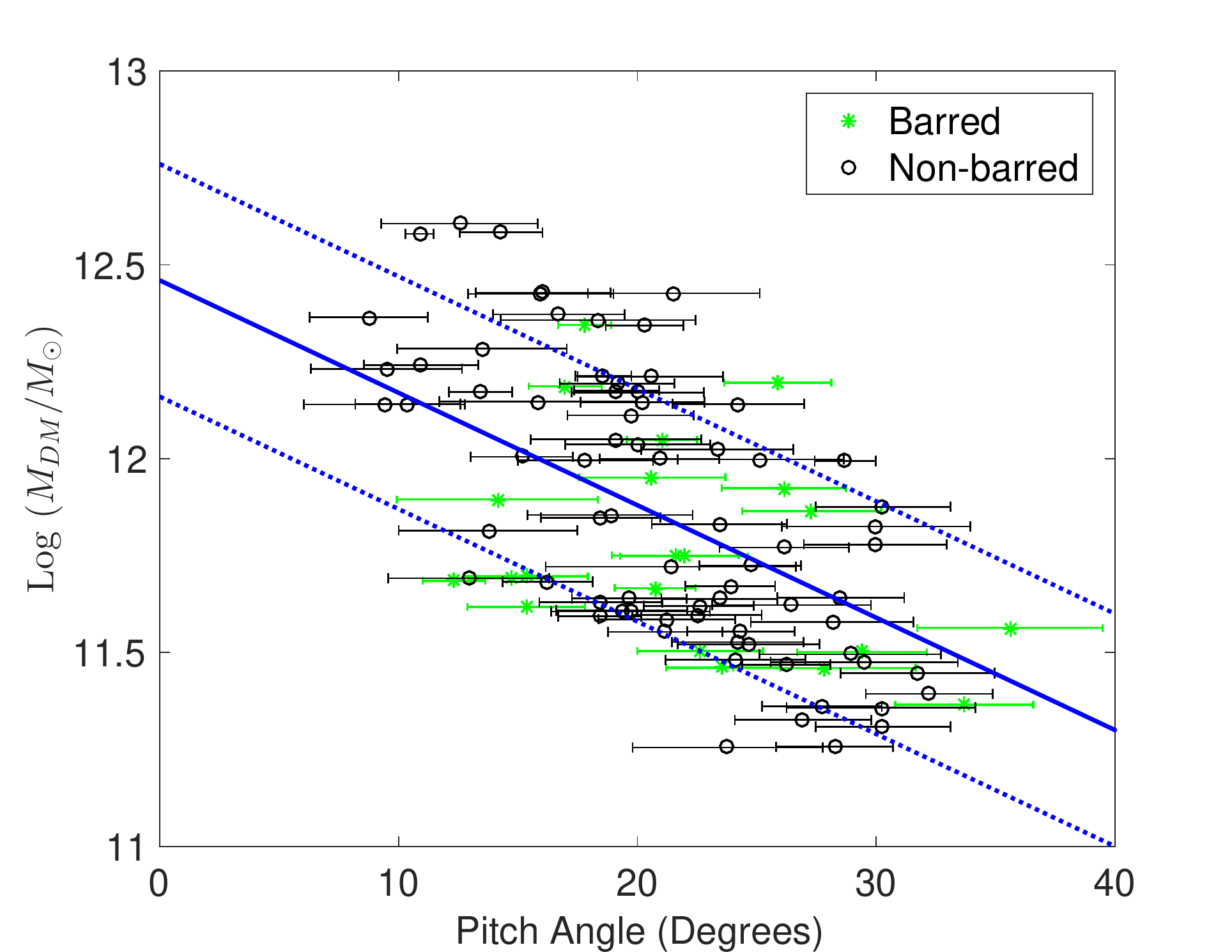}
  \includegraphics[width=\columnwidth]{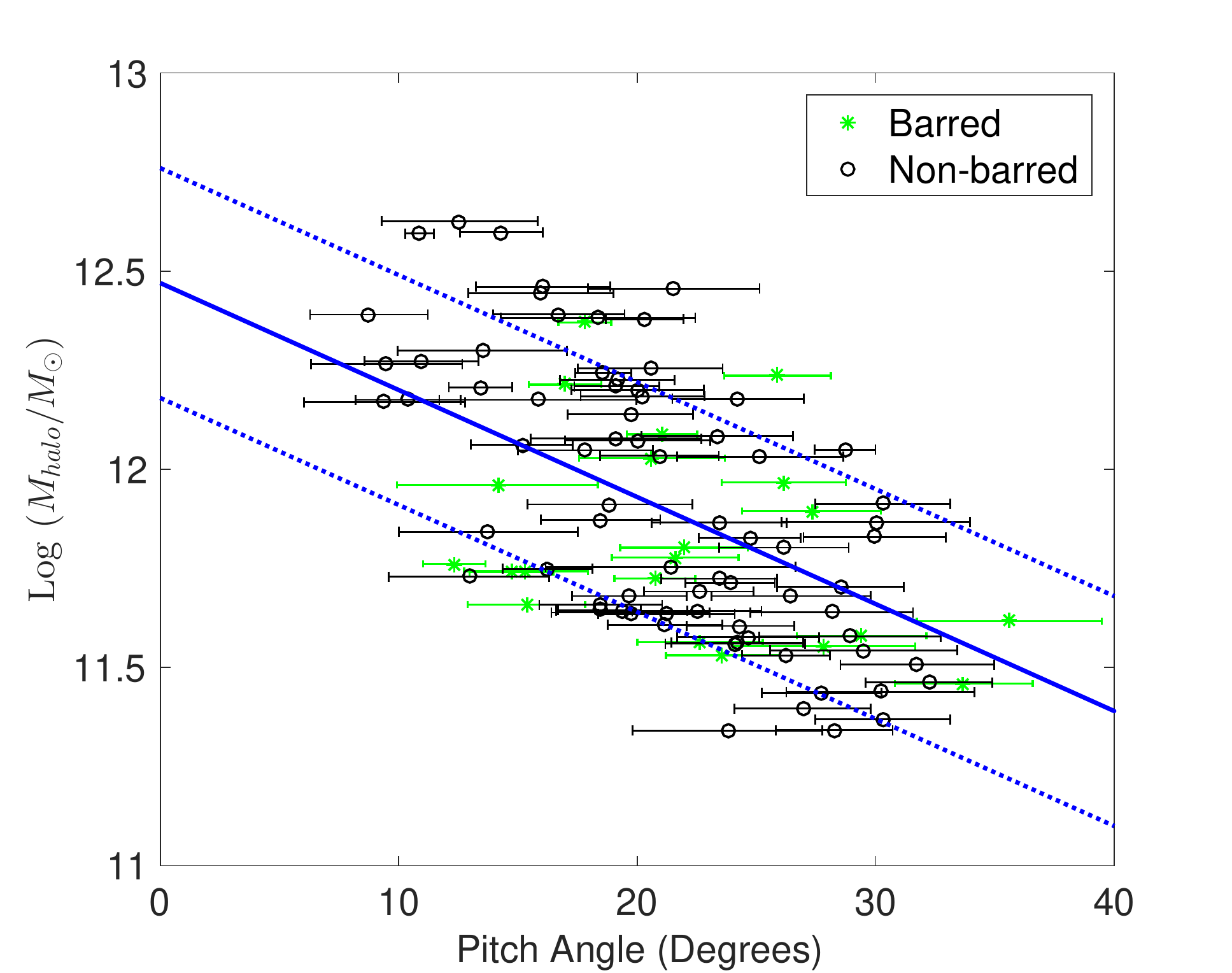}
  \caption{The Illustris predictions for the DM mass and halo mass relative to pitch angle for our spiral sample are shown by the solid blue lines, which represent the best fits to 
the data: left is the $M_{DM}-P$ relation and right is the $M_{halo}-P$ relation. Galaxies with bars are shown by green stars. Galaxies with no clear sign of a bar are shown by black circles. The total 
RMS scatters are plotted as dotted blue lines above and below the best-fit lines: $\pm0.30$ dex in the $\log(M_{DM}/M_{\sun})$ direction and $\pm0.29$ dex in the $\log(M_{halo}/M_{\sun})$ direction. 
Note that the Illustris predictions for these correlations are tighter than the one between pitch angle and the BH mass.}
  \label{FIGURE-6}
\end{figure*}

Based on limited data, \citet{Seigar2014} uncovered a weak trend between pitch angle (used as a proxy for the BH mass) and DM concentration such that galaxies with tightly wound arms have 
particularly high concentrations. Furthermore, cosmological simulations often reveal an anti-correlation between DM mass and concentration \citep[e.g.,][]{Bullock2001,vandenBosch2014}.
Here, we show that the Illustris simulation results in an anti-correlation between DM mass and pitch angle for our spiral sample (see Figure 6). In conjunction with the DM mass 
versus concentration anti-correlation \citep{Bullock2001,vandenBosch2014} and a DM mass versus pitch angle correlation (Figure 6), we expect to derive a positive correlation between 
pitch angle and DM concentration. This is opposite to the weak trend presented in \citet{Seigar2014}. However, given the large scatter in the DM mass versus concentration relations revealed by 
the simulations, and the limitations of the observational data discussed by \citet{Seigar2014} no strong conclusions can be made. Indeed, \citet{Seigar2014} commented that more observational data 
is needed to really test the predictions made in theoretical studies, such that it would be possible to truly determine, observationally, how DM mass and pitch angle are related or, 
indeed, if they are related. Moreover, it would be worth testing this link with the future Illustris dark matter halo catalogs \citep{Chua2016} in a future study.

Many works \citep[e.g.,][]{Graham2008,Hu2008,Gultekin2009,Graham2013} have identified an offset in the supermassive black hole$-$bulge relations between galaxies with barred and 
non-barred morphologies. For the pitch angle scaling relations, we find no offset between the barred (green stars) and non-barred (black circles) galaxies of our spiral sample in Illustris.

\section{Black Hole Scaling Relations Beyond the Bulge}

In this section, we focus on the $z=0$ Illustris central galaxies with $10.0<\log(M_{\star,2R}/M_{\sun})<13.0$ to study the BH mass scaling relations beyond the bulge, i.e., the $M_{BH}-M_{\star,total}$, the $M_{BH}-M_{DM}$ and $M_{BH}-M_{halo}$ relations. We adopt $M_{\star,total}$ directly from the simulation outputs as the total mass of the stellar particles bound to the subhalo.
Here, we only investigate central galaxies, i.e., the main halo of each FoF group, in order to avoid satellite galaxies for which the halo mass is poorly defined and much of the mass may be assigned to the parent halo. In our spiral sample, there are 81 central galaxies: 17 barred, 64 non-barred. In this section, we only display these central spiral galaxies. Note that \citet{Snyder2015} also presented these relations as a function of galaxy morphology. Here, we investigate these predictions by comparing them with other theoretical results and observational constraints at $z=0$. The aim is to clearly understand these different relations, e.g., which one may be the strongest correlation, whether different galaxy types exhibit weaker/stronger physical links with their BHs. Moreover, a better understanding of the relations beyond the bulge could provide a benchmark for high-redshift studies which cannot avail
themselves of bulge masses or dynamical BH masses \citep{ReinesVolonteri2015}.

\subsection{The $M_{BH}-M_{\star,total}$ Relation}
In Figure 7, we display the Illustris prediction of the $M_{BH}-M_{\star,total}$ relation for central galaxies with the solid yellow line, which presents the best fit to the simulation data: 
$\log(M_{BH}/M_{\sun})=(1.53\pm0.02)\log(M_{\star,total}/M_{\sun})-(8.85\pm0.26)$, with R$^{2}=0.75$, RMSE $=0.36$. For comparison, we have also plotted the observed 
$M_{BH}-M_{\star,total}$ relations by \citet{ReinesVolonteri2015}, which used a sample of 262 local broad-line AGN and 79 galaxies with dynamical BH masses.
Their total stellar mass measurements rely on mass-to-light ratios. The authors defined a relation for AGN host galaxies (solid blue line) that has a similar slope to early-type 
galaxies with quiescent BHs (dashed blue line), but a normalization that is more than an order of magnitude lower. They also suggested that a potential origin for these different relations could
be differences in host galaxy properties (e.g., the Hubble types), claiming that spirals/discs tend to overlap with AGN host galaxies.
Although the Illustris galaxies lie within the observational relations of \citet{ReinesVolonteri2015}, it seems that Illustris under-estimates the BH masses of early-type 
galaxies for a given $M_{\star,total}$ when compared to the observational relation (dashed blue line). Moreover, Illustris appears to favor a single linear-relation for all galaxy types, disagreeing
with the findings of \citet{ReinesVolonteri2015}. To explore the relation based on bar morphology, the barred and non-barred spiral galaxies in our sample are depicted by different 
signs (green dots and red crosses, respectively). The dashed yellow line is the best fit to our spiral sample, which is consistent with the best fit that is obtained from all galaxy types. 
Note that the apparent scarcity of the galaxies with $M_{\star,total} \sim 10^{10.5} M_{\sun}$ in our spiral sample is due to the high abundance of the peculiar galaxies, i.e. ringed galaxies, in 
this mass range \citep{Snyder2015}. We also compare with the prediction of the high-resolution cosmological simulation $MassiveBlackII$ \citep{DeGraf2015}, showing remarkable agreement 
with the Illustris prediction with a slightly shallow slope (see the magenta line). Their relation, which is derived for all galaxy types, is in remarkable agreement with our best-fit for 
our spiral sample. This supports the idea that BHs in spiral galaxies correlate with $M_{*,total}$ in a similar way to those in early-type galaxies do. \citet{DeGraf2015} showed 
that the low-end of their $M_{BH}-\sigma_{sph}$ relation tends to lie above the observations, which is consistent with \citet{Sijacki2015}, who found similar behaviour at the low-end. 
However, the authors showed that their $M_{BH}-M_{\star,total}$ relation shows good agreement with the observations at the high-end as well as at the low-end of the relation. Based on the 
consistent predictions between Illustris and $MassiveBlackII$, this also implies good agreement between Illustris and the observations for both ends of the relation. 

\begin{figure*}
  \includegraphics[width=13cm]{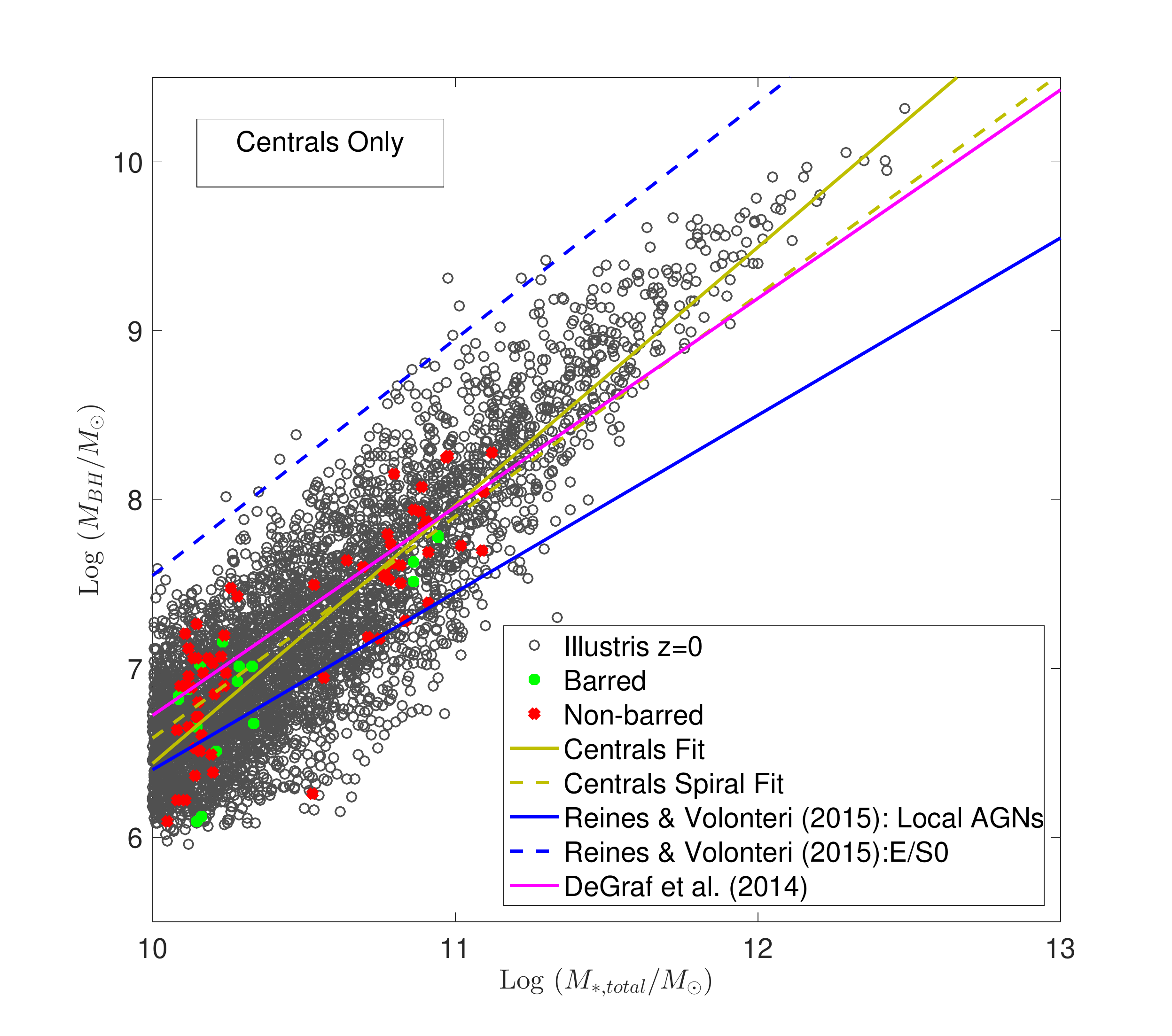}
  \caption{The Illustris prediction of the $M_{BH}-M_{\star,total}$ relation for central galaxies is shown by the solid yellow line, which represents the best fit to the simulated data: the slope and normalization are $1.53\pm0.02$ and $-8.85\pm0.26$ (R$^{2}=0.75$, RMSE$=0.36$) for the $\log(M_{BH}/M_{\sun})-\log(M_{\star,total}/M_{\sun})$ relation. While the red crosses indicate our non-barred spiral sample from Illustris, the green filled circles indicate barred ones. The best-fit to our spiral sample is shown by the dashed yellow line. For comparison, we plot the observed 
$M_{BH}-M_{\star,total}$ relations by \citet{ReinesVolonteri2015}, represented by the solid blue line for local AGNs and the dashed blue line for E/S0 galaxies. These authors also suggested that 
spirals and discs follow the same relation with local AGNs. We also compare with the relation by \citet{DeGraf2015} from the high-resolution cosmological simulation $MassiveBlackII$, represented
by the magenta line.}
  \label{FIGURE-7}
\end{figure*}

\subsection{The $M_{BH}-M_{DM}$ Relation}
Figure 8 shows the Illustris prediction of the $M_{BH}-M_{DM}$ relation for central galaxies. The best fit to the simulation data is 
$\log(M_{BH}/M_{\sun})=(1.55\pm0.02)\log(M_{DM}/M_{\sun})-(11.26\pm0.20)$, with R$^{2}=0.88$, RMSE $=0.25$. The best fit to our simulated spiral sample 
is depicted by the dashed yellow line, which is very close to the one for all types (solid yellow line). Both the barred (green filled circles) and non-barred (red crosses) spirals follow 
the mean relation closely without falling systematically above or below the relation. 

\begin{figure*}
  \includegraphics[width=13cm]{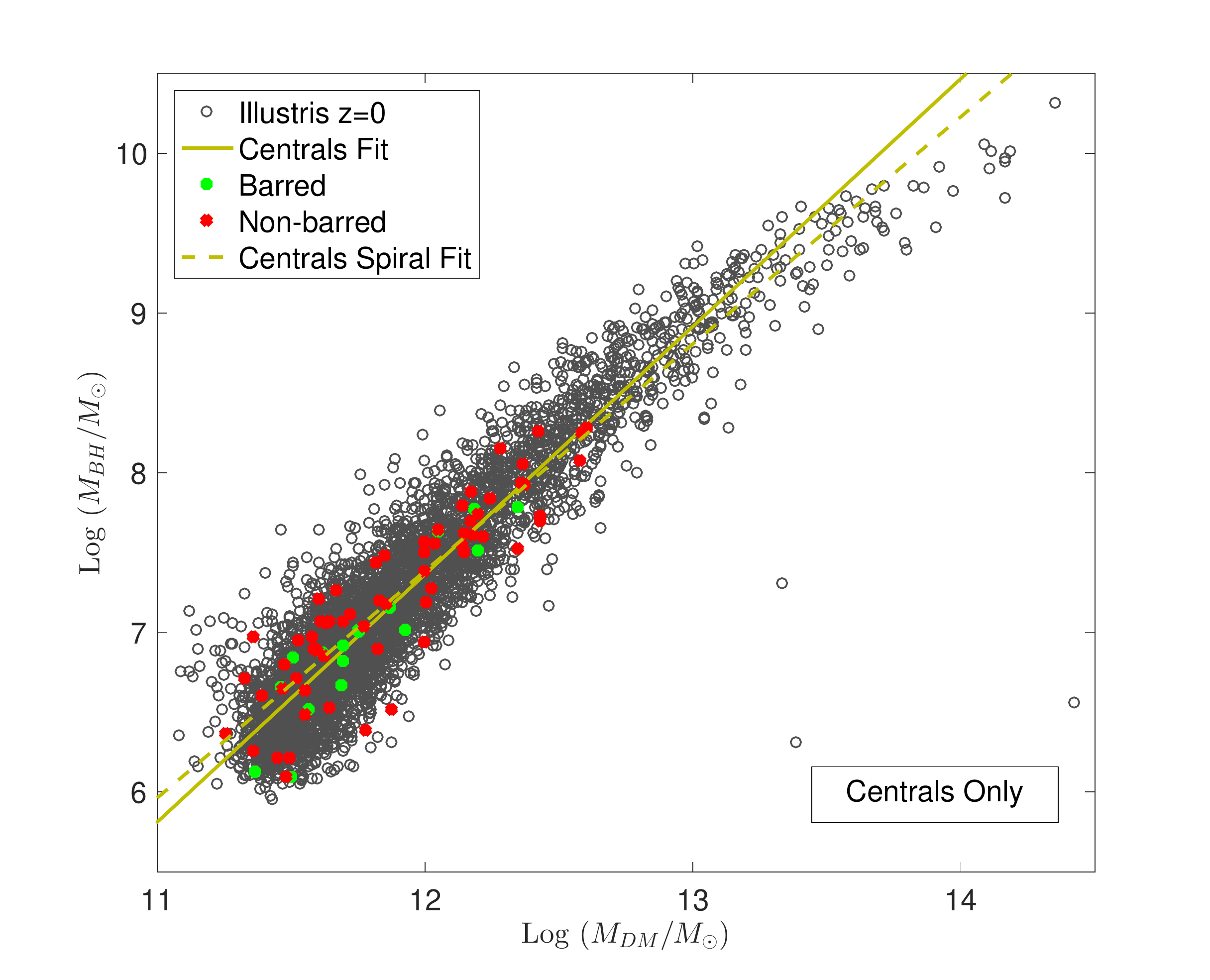}
  \caption{The Illustris prediction of the $M_{BH}-M_{DM}$ relation for central galaxies is shown by the solid yellow line, which represents the best fit to the simulated data: 
$\log(M_{BH}/M_{\sun})=(1.55\pm0.02)*\log(M_{DM}/M_{\sun})-(11.26\pm0.20)$ (R$^{2}=0.88$, RMSE$=0.25$). The dashed yellow line represents the best fit to our spiral sample, which is shown 
by the red crosses (non-barred) and the filled green circles (barred).}
  \label{FIGURE-8}
\end{figure*}

\subsection{The $M_{BH}-M_{halo}$ Relation}
Here, we investigate the $M_{BH}-M_{halo}$ relation for central galaxies, where we adopt $M_{halo}$ as the mass enclosed within a sphere, centered on the potential minimum of the halo,
that has a mean internal density of 200 times the critical density of the Universe. 
Figure 9 compares the Illustris prediction of the $M_{BH}-M_{halo}$ relation with other theoretical results and the observational constraints. Most self-regulating theoretical models of 
galactic physics predict a fundamental connection between the central BH mass and the total mass of the host galaxy \citep{Adams2001,ElZant2003,Haehnelt1998,Monaco2000,SilkRees1998}. 
The galaxy models, which study the interaction between the DM halos of galaxies and baryonic matter, predict that halo properties determine the masses of bulge and central
BH \citep{Cattaneo2001,ElZant2003,Hopkins2005}. However, there are inconsistencies in the observational studies: while some studies argued for \citep{Seigar2011,Volonteri2011},
some argued against \citep{KormendyBender2011} a coupling of the BH mass and the properties of the DM halo. It is not an easy task to investigate this relation observationally due 
to the uncertainties in measurement of the halo mass. Some of the first indirect observational evidence for the existence of a $M_{BH}-M_{halo}$ relation come from the study of 
\citet{Ferrarese2002}, which first derived a correlation between $\sigma_{sph}$ and the galaxy's circular velocity ($v_{c}$) for a sample of 20 elliptical galaxies and 16 spiral galaxies, 
then translated it to a $M_{BH}-M_{halo}$ relation. While using the $M_{BH}-\sigma_{sph}$ relation \citep{Ferrarese2000} to estimate the BH mass, the author demonstrated the effect of the method 
used to connect $v_{c}$ to $M_{halo}$ via $v_{vir}$. The solid magenta line shows the best-fit $M_{BH}-M_{halo}$ relation obtained by \citet{Ferrarese2002} using the cosmological prescription 
of \citet{Bullock2001} to relate $v_{c}$ and $v_{vir}$. The dashed magenta line shows the resulting relation if $v_{vir}/v_{c}=1$ is assumed while the dotted magenta line presents 
the resulting relation if $v_{vir}/v_{c}=1.8$ is used, as proposed by \citet{Seljak2002}. Among the relations obtained by \citet{Ferrarese2002}, the Illustris prediction agrees better with 
the solid magenta line, i.e., the one derived from the cosmological prescription of \citet{Bullock2001}. The solid cyan line is the best fit for Illustris central galaxies from \citet{Snyder2015}, 
while the solid green line is our best fit: $\log(M_{BH}/M_{\sun})=(1.62\pm0.02)*\log(M_{halo}/M_{\sun})-(12.07\pm0.23)$ (R$^{2}=0.85$, RMSE$=0.28$). 
Note that \citet{Snyder2015} fitted the Illustris data by using orthogonal distance regression method \citep{BoggsRogers1990} while we used bisquare weights. 
Later, \citet{Bandara2009} investigated this relation by using gravitational lens modelling to determine the total mass. They estimated the BH masses from the $M_{BH}-\sigma_{sph}$ 
relation by \citet{Gultekin2009}. They recovered a non-linear correlation between $M_{BH}$ and $M_{halo}$, suggesting a slope of $\sim1.67$ that implies a 
merger-driven, feed-back regulated process for the growth of supermassive black holes. Based on their argument, the \citet{Snyder2015} fit with a slope of $\sim1.69$ implies a merger-driven, 
feed-back regulated process, and supports the conclusion of \citet{Sijacki2015} that both BH feedback and BH$-$BH mergers are important ingredients to produce the $M_{BH}-M_{sph}$ relation. 
Using self-consistent simulations of the co-evolution of the BH and galaxy populations, \citet{BoothSchaye2010} demonstrated a very good agreement between the observational 
determination of \citet{Bandara2009} and their simulation when the simulation was only tuned to match the normalization of the relations between $M_{BH}$ and the galaxy stellar properties. 
The mean Illustris trend seems to differ from those of \citet{BoothSchaye2010} and \citet{Bandara2009} by several standard deviations, but stays within the region defined 
by \citet{Ferrarese2000}. In addition, \citet{BoothSchaye2010} suggested a primary link to halo binding energy rather than the halo mass, with the central halo concentration playing a role in the relation's scatter. It would be worth investigating this link with the future Illustris DM halo data catalogs \citep{Chua2016} in a future study. 

In addition, the best fit to our spiral sample, shown by the solid yellow line in Figure 9, is consistent with the best fits to all galaxy types (solid cyan, green and magenta lines), implying that
spirals in Illustris follow the mean relation very closely. This suggests the existence of a tight correlation between the galactic properties of star-forming, 
blue galaxies and their BHs. Also, we do not observe any systematic behaviour based on bar morphology such as being well above or below the mean relation, suggesting the presence 
or absence of a bar does not affect the link between the BH and its host galaxy significantly. Moreover, we highlight that the simulated galaxies present a smaller scatter around 
the mean $M_{BH}-M_{halo}$ relation when compared to the mean $M_{BH}-M_{\star, total}$ relation.

\begin{figure*}
  \includegraphics[width=13cm]{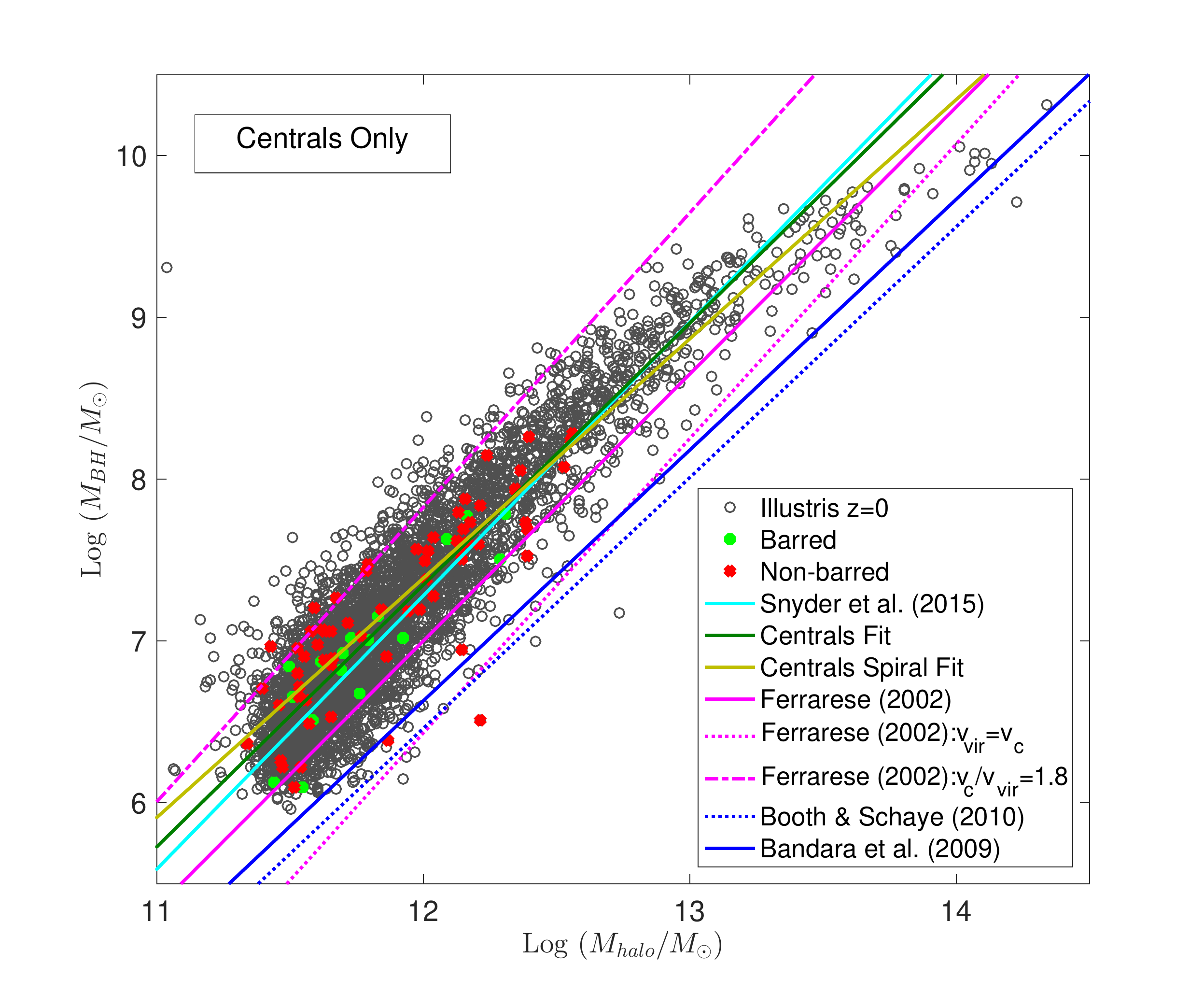}
  \caption{The Illustris prediction of the $M_{BH}-M_{halo}$ relation for central galaxies is compared with other theoretical predictions and the local observations. 
The solid cyan line is the best fit from \citet{Snyder2015} while the solid green line is our best fit for central galaxies. The solid yellow line 
is the best fit to our spiral sample, which is shown by the red crosses (non-barred) and green filled circles (barred). The magenta lines are same as in Figure 6 of 
\citet{Ferrarese2002}: the solid line corresponds to the best fit using the prescription of \citet{Bullock2001} to relate $v_{vir}$ to $v_{c}$, the dot-dashed 
line shows where the galaxies would lie if $v_{vir}=v_{c}$, and the dotted line shows where they would move to if $v_{c}/v_{vir}=1.8$, as proposed by \citet{Seljak2002}. 
The solid blue line shows the observational determination of \citet{Bandara2009} while the dotted blue line shows the result from the simulation of \citet{BoothSchaye2010}.}
  \label{FIGURE-9}
\end{figure*}

\section{Conclusions}
We investigate the supermassive black hole$-$galaxy connection beyond the bulge in the high-resolution cosmological simulation Illustris. Our work is complementary to \citet{Sijacki2015}, 
which studied the black hole$-$bulge scaling relations as a function of galaxy morphology, color and specific star formation rate. Here we study these predictions by comparing 
with other theoretical results and observational constraints at $z=0$. 

First, we use a randomly selected sample of Illustris spiral galaxies that do not have rings, and study the spiral arm morphology in multiple wavebands. 
We present the existence of a 1:1 relation between the $B$- and $K$-band pitch angles of our sample in Illustris, which is consistent with the observational 
findings of \citet{Seigar2006} and \citet{Davis2012}. Then, we explore the Illustris predictions for the BH mass, DM mass and halo mass relative to pitch angle 
for our spiral sample. We derive an Illustris prediction of the $M_{BH}-P$ relation, that is consistent with observational constraints. More importantly, we 
obtain a tighter correlation between pitch angle and DM/halo mass. Based on the Illustris predictions, we conclude that pitch angle can be used as a tracer for DM/halo mass.

In addition, we focus on the $z=0$ Illustris central galaxies with $10.0<\log(M_{\star,2R}/M_{\sun})<13.0$ to study the $M_{BH}-M_{\star,total}$, the $M_{BH}-M_{DM}$ and $M_{BH}-M_{halo}$ 
relations by comparing with other theoretical studies and observational constraints. We find that Illustris establishes very tight correlations between the BH mass and large-scale 
properties of the host galaxy, not only for early-type galaxies but also low-mass, blue and star-forming galaxies, regardless of bar morphology. The tight relations shown in this 
work are strongly suggestive that halo properties play an important role in determining those of the galaxy and its supermassive black hole.

\section*{Acknowledgements}
BMP and MSS wish to thank the generous support of the University of Minnesota, Duluth and the Fund for Astrophysical Research. We also wish to thank the anonymous referee whose comments
greatly improved the content of this paper. We also thank Benjamin L. Davis for providing us the pitch angle measurements of \citet{Berrier2013}.

This research has made use of the Illustris database. 
The Illustris project acknowledges support from many sources: support by the DFG Research Centre SFB-881 "The Milky Way System" through project A1, and by the European Research Council 
under ERC-StG EXAGAL-308037, support from the HST grants program, number HST-AR- 12856.01-A, support for program $\#$12856 by NASA through a grant from the Space Telescope Science Institute, 
which is operated by the Association of Universities for Research in Astronomy, Inc., under NASA contract NAS 5-26555, support from NASA grant NNX12AC67G and NSF grant AST-1312095, support 
from the Alexander von Humboldt Foundation, NSF grant AST-0907969, support from XSEDE grant AST-130032, which is supported by National Science Foundation grant number OCI-1053575. 
The Illustris simulation was run on the CURIE supercomputer at CEA/France as part of PRACE project RA0844, and the SuperMUC computer at the Leibniz Computing Centre, Germany, 
as part of project pr85je. Further simulations were run on the Harvard Odyssey and CfA/ITC clusters, the Ranger and Stampede supercomputers at the Texas Advanced Computing Center 
through XSEDE, and the Kraken supercomputer at Oak Ridge National Laboratory through XSEDE.

\begin{table*}
	\centering
	\caption{Our spiral galaxy sample. Columns: (1) Illustris subhalo ID. (2) Bar morphology: y for barred, n for non-barred. (3) Harmonic mode. (4) Highest amplitude of the chosen harmonic mode in the $B$-band. (5) $B$-band pitch angle in degrees. (6) $R$-band pitch angle in degrees. (7) $K$-band pitch angle in degrees. (8) $M_{BH}$ is the BH mass contained within the stellar half-mass radius, retrieved from the Illustris database. (9)$M_{*,total}$ is the total stellar mass bound to the subhalo, retrieved from the Illustris database. (10) $M_{DM}$ is the mass of dark matter bound to the subhalo, retrieved from the Illustris database. (11) $M_{halo}^{(2)}$, which is used in Section 2, is the total mass of all particles (all types) bound to the subhalo, retrieved from the Illustris database. (12) $M_{halo}^{(3)}$, which is used for central galaxies in Section 3, is the mass enclosed within a sphere, centered on the potential minimum of the halo, that has a mean internal density of 200 times the critical density of the Universe.}
	\label{tab:1}
	\scriptsize
	\begin{tabular}{lccccccccccc}  
          \hline
          {ID} & Bar & m & $A(p_{max},m)$ & {P$_{B}$} & {P$_{R}$} & {P$_{K}$}  & {$\log(\frac{M_{BH}}{M_{\sun}})$} & {$\log(\frac{M_{*,total}}{M_{\sun}})$} & $\log(\frac{M_{DM}}{M_{\sun}})$ & $\log(\frac{M_{halo}^{(2)}}{M_{\sun}})$ & $\log(\frac{M_{halo}^{(3)}}{M_{\sun}})$\\
           (1) & (2) &   (3)    &    (4)   &     (5)    &         (6)                &          (7)                   & (8)                      &    (9) & (10) & (11) & (12) \\        
\hline
41098&	y& 2& 0.17&	$15.61\pm3.06$& $14.65\pm4.59$ & $14.14\pm4.21$ &	8.08 &	11.07 &	11.89 &	11.96 & $-$	\\
66096&	y& 2& 0.23&	$24.72\pm3.44$& $22.44\pm3.53$ & $27.84\pm3.82$ &	6.49 &	10.18 &	11.46 &	11.55 &	$-$	\\
167869&	n& 3& 0.05&	$12.68\pm1.12$& $9.60\pm1.79$ & $16.24\pm1.88$ &	7.61 &	10.82 &	11.68 &	11.75 &	$-$	\\
183689&	n& 6& 0.04&	$22.70\pm1.65$& $22.34\pm2.19$ & $22.58\pm2.30$ &	7.31 &	10.77 &	11.62 &	11.69 &	$-$	\\
250637&	n& 3& 0.04&	$19.82\pm3.64$& $21.67\pm1.74$ & $23.43\pm2.43$ &	7.22 &	10.82 &	11.64 &	11.72 & $-$	\\
274690&	y& 6& 0.08&	$21.53\pm3.89$& $22.32\pm3.04$ & $20.74\pm1.69$ &	7.04 &	10.30 &	11.67 &	11.72 & $-$	\\
278699&	n& 2& 0.08&	$21.37\pm3.65$& $22.72\pm2.75$ & $24.72\pm2.13$ &	7.83 &	11.10 &	11.72 &	11.83 &	$-$	\\
287939&	n& 2& 0.12&	$26.53\pm3.43$& $26.06\pm4.42$ & $30.29\pm2.83$ &	7.14 &	10.18 &	11.31 &	11.37 &	$-$	\\
292278&	n& 2& 0.20&	$13.74\pm3.78$& $13.24\pm3.17$ & $10.39\pm2.20$ &	8.06 &	10.97 &	12.14 &	12.18 &	$-$	\\
293192&	n& 3& 0.07&	$9.25\pm2.82$	& $9.22\pm1.81$ & $9.49\pm3.17$ &	7.84 &	11.06 &	12.23 &	12.27 &	$-$	\\
295983&	n& 3& 0.04&	$28.15\pm3.38$& $27.00\pm2.38$ & $23.79\pm3.98$ &	7.12 &	10.24 &	11.25 &	11.34 & $-$	\\
308471&	n& 5& 0.05&	$20.18\pm1.00$& $19.96\pm1.26$ & $19.71\pm2.64$ &	7.76 &	10.69 &	12.11 &	12.14 & $-$	\\
315319&	y& 2& 0.20&	$15.87\pm1.24$& $16.18\pm1.11$ & $20.62\pm3.06$ &	7.46 &	10.96 &	11.95 &	12.03 &	$-$	\\
318403&	n& 3& 0.07&	$18.89\pm3.65$& $16.76\pm4.38$ & $20.54\pm3.04$ &	7.64 &	11.01 &	12.21 &	12.25 &	$-$	\\
339972&	n& 3& 0.03&	$14.94\pm1.39$& $14.41\pm1.47$ & $12.55\pm3.27$ &	8.28 &	11.12 &	12.61 &	12.63 & 12.56	\\
348613&	n& 3& 0.04&	$11.56\pm2.99$& $10.51\pm3.22$ & $14.30\pm1.73$ &	8.24 &	10.97 &	12.58 &	12.60 &	12.55	\\
349838&	n& 3& 0.03&	$10.86\pm1.88$& $10.76\pm0.77$ & $10.88\pm0.60$ &	8.07 &	10.89 &	12.58 &	12.60 &	12.53	\\
353567&	n& 2& 0.24&	$16.52\pm1.92$& $16.67\pm2.25$ & $16.05\pm2.82$ &	7.69 &	11.09 &	12.43 &	12.46 &	12.39	\\
357033&	n& 3& 0.09&	$18.40\pm1.88$& $19.08\pm1.35$ & $18.35\pm4.08$ &	7.94 &	10.86 &	12.36 &	12.38 &	12.35	\\
360070&	n& 5& 0.07&	$24.01\pm1.45$& $24.32\pm2.82$ & $21.53\pm3.60$ &	7.73 &	11.02 &	12.43 &	12.46 &	12.38	\\
365003&	n& 6& 0.04&	$19.67\pm1.60$& $20.43\pm1.46$ & $20.31\pm1.63$ &	7.53 &	10.78 &	12.34 &	12.38 &	12.39	\\
373991&	n& 3& 0.08&	$13.92\pm3.32$& $17.14\pm3.31$ & $15.96\pm3.04$ &	8.26 &	10.98 &	12.43 &	12.44 &	12.39	\\
376968&	n& 3& 0.04&	$9.58\pm1.89$	& $9.42\pm1.69$ & $8.75\pm2.47$ &	8.05 &	11.09 &	12.36 &	12.39 &	12.36	\\
377089&	y& 3& 0.12&	$19.07\pm3.58$& $21.96\pm2.26$ & $25.88\pm2.24$ &	7.51 &	10.86 &	12.20 &	12.24 &	12.29	\\
379786&	y& 6& 0.14&	$15.92\pm2.21$& $18.31\pm1.45$ & $17.80\pm1.11$ &	7.79 &	10.94 &	12.34 &	12.37 &	12.31	\\
381137&	n& 4& 0.05&	$19.15\pm2.90$& $18.24\pm2.67$ & $16.71\pm2.76$ &	7.93 &	10.88 &	12.37 &	12.39 &	12.35	\\
390346&	n& 4& 0.08&	$19.07\pm2.32$& $18.62\pm2.74$ & $19.15\pm2.40$ &	7.74 &	10.79 &	12.19 &	12.23 &	12.18	\\
393605&	n& 4& 0.06&	$21.11\pm1.41$& $22.38\pm1.74$ & $20.22\pm2.60$ &	7.50 &	10.82 &	12.15 &	12.18 &	12.14	\\
393984&	n& 2& 0.10&	$11.95\pm1.23$& $11.35\pm1.82$ & $10.95\pm2.39$ &	7.84 &	10.90 &	12.24 &	12.27 &	12.21	\\
396937&	n& 3& 0.04&	$14.52\pm2.82$& $15.31\pm2.89$ & $13.50\pm3.55$ &	8.15 &	10.80 &	12.28 &	12.30 &	12.24	\\
397204&	n& 3& 0.05&	$19.09\pm3.11$& $18.50\pm2.10$ & $18.57\pm1.17$ &	7.60 &	10.70 &	12.21 &	12.24 &	12.21	\\
397695&	n& 2& 0.08&	$19.44\pm2.16$& $21.75\pm2.57$ & $9.41\pm3.38$ &	7.79 &	10.77 &	12.14 &	12.17 &	12.13	\\
400083&	n& 2& 0.13&	$26.10\pm3.28$& $27.10\pm2.79$ & $28.71\pm1.28$ &	6.94 &	10.56 &	11.99 &	12.05 &	12.14	\\
401370&	n& 2& 0.08& 	$20.52\pm1.47$& $21.07\pm1.68$ & $19.14\pm1.78$ &	7.61 &	10.82 &	12.18 &	12.21 &	12.16	\\
401474&	y& 4& 0.05&	$15.78\pm2.38$& $17.23\pm2.70$ & $16.98\pm1.52$ &	7.78 &	10.94 &	12.19 &	12.21 &	12.17	\\
401856&	n& 2& 0.10&	$26.53\pm3.43$& $26.06\pm4.42$ & $30.29\pm2.83$ &	6.51 &	10.16 &	11.88 &	11.91 &	12.21	\\
403323&	n& 3& 0.12&	$14.09\pm1.46$& $13.98\pm1.67$ & $13.43\pm1.33$ &	7.88 &	10.91 &	12.17 &	12.21 &	12.16	\\
403642&	n& 2& 0.11&	$20.34\pm3.03$& $19.82\pm4.07$ & $15.83\pm4.13$ &	7.62 &	10.79 &	12.15 &	12.18 &	12.12	\\
406286&	n& 3& 0.06&	$20.68\pm1.72$& $20.73\pm2.51$ & $20.02\pm2.77$ &	7.69 &	10.91 &	12.17 &	12.20 &	12.15	\\
407890&	n& 3& 0.21&	$16.89\pm2.51$& $19.20\pm2.62$ & $24.23\pm2.76$ &	7.55 &	10.77 &	12.14 &	12.18 &	12.14	\\
408954&	n& 3& 0.10&	$20.02\pm3.13$& $18.23\pm3.59$ & $18.86\pm3.45$ &	7.18 &	10.75 &	11.85 &	11.91 &	11.94	\\
412343&	n& 2& 0.19&	$23.86\pm3.66$& $21.81\pm2.31$ & $15.16\pm2.14$ &	7.19 &	10.71 &	12.00 &	12.06 &	11.99	\\
412515&	n& 4& 0.09&	$27.73\pm3.68$& $28.19\pm3.88$ & $23.35\pm3.18$ &	7.28 &	10.84 &	12.02 &	12.08 &	12.04	\\
416531&	y& 4& 0.16&	$24.36\pm2.73$& $22.53\pm1.43$ & $21.04\pm1.47$ &	7.63 &	10.86 &	12.05 &	12.09 &	12.09	\\
417495&	n& 3& 0.07&	$15.75\pm3.14$& $17.46\pm2.28$ & $20.01\pm3.04$ &	7.56 &	10.77 &	12.04 &	12.07 &	12.02	\\
417736&	n& 4& 0.05&	$20.52\pm1.36$& $20.96\pm2.05$ & $17.83\pm2.83$ &	7.39 &	10.91 &	12.00 &	12.05 &	12.01	\\
420546&	n& 4& 0.09&	$18.29\pm3.48$& $17.77\pm3.49$ & $19.11\pm3.58$ &	7.64 &	10.64 &	12.05 &	12.08 &	12.04 	\\
429333&	n& 4& 0.03&	$16.55\pm3.21$& $16.42\pm2.11$ & $20.93\pm2.49$ &	7.57 &	10.77 &	12.00 &	12.03 &	11.98	\\
432924&	n& 4& 0.09&	$20.51\pm2.85$& $11.44\pm1.81$ & $25.16\pm3.48$ &	7.50 &	10.54 &	12.00 &	12.03 &	12.01	\\
437478&	y& 2& 0.27&	$23.92\pm3.33$& $24.59\pm3.15$ & $26.14\pm2.60$ &	7.02 &	10.33 &	11.92 &	11.97 &	11.93	\\
437661&	n& 3& 0.03&	$25.66\pm2.72$& $25.34\pm2.39$ & $29.96\pm2.99$ &	6.39 &	10.20 &	11.78 &	11.83 &	11.87	\\
445191&	n& 5& 0.03&	$34.62\pm2.89$& $34.82\pm3.15$ & $30.00\pm3.95$ &	6.90 &	10.24 &	11.83 &	11.87 &	11.86	\\
450471&	y& 3& 0.22&	$19.62\pm3.12$& $13.49\pm1.19$ & $12.33\pm1.31$ &	6.67 &	10.34 &	11.69 &	11.76 &	11.76	\\
450907&	y& 5& 0.08&	$26.75\pm1.91$& $24.89\pm3.32$ & $27.30\pm2.91$ &	7.15 &	10.23 &	11.86 &	11.89 &	11.83	\\
453583&	n& 4& 0.07&	$18.43\pm1.51$& $18.31\pm1.75$ & $21.23\pm2.86$ &	6.90 &	10.11 &	11.59 &	11.63 &	11.56	\\
458148&	n& 2& 0.14&	$16.86\pm2.23$& $18.38\pm2.79$ & $23.43\pm2.83$ &	7.20 &	10.24 &	11.83 &	11.87 &	11.84	\\
461309&	n& 3& 0.06&	$20.93\pm2.86$& $22.48\pm2.71$ & $18.47\pm2.50$ &	7.48 &	10.26 &	11.85 &	11.87 & 11.79	\\
461677&	y& 2& 0.16&	$16.62\pm3.31$& $14.96\pm2.63$ & $21.96\pm2.68$ &	7.01 &	10.28 &	11.75 &	11.80 & 11.79	\\
466545&	n& 2& 0.12& 	$10.37\pm3.19$& $13.03\pm2.08$ & $13.75\pm3.74$ &	7.43 &	10.28 &	11.81 &	11.84 &	11.79	\\
468379&	n& 3& 0.18&	$23.72\pm2.86$ & $25.19\pm2.52$ & $26.15\pm2.71$ &	7.03 &	10.20 &	11.77 &	11.80 &	11.77	\\
468683&	n& 4& 0.21&	$24.03\pm2.38$ & $23.42\pm1.07$ & $28.52\pm2.65$ &	6.53 &	10.14 &	11.64 &	11.70 &	11.65	\\
\hline
\end{tabular}
\end{table*} 

\begin{table*}
	\centering
	\contcaption{Our spiral galaxy sample}
	\label{tab:1}
        \scriptsize
	\begin{tabular}{lccccccccccc}  
          \hline
          {ID} & Bar &    m     & $A(p_{max},m)$ & {P$_{B}$} & {P$_{R}$} & {P$_{K}$}  & {$\log(\frac{M_{BH}}{M_{\sun}})$} & {$\log(\frac{M_{*,total}}{M_{\sun}})$} & $\log(\frac{M_{DM}}{M_{\sun}})$ & 
$\log(\frac{M_{halo}^{(2)}}{M_{\sun}})$ & $\log(\frac{M_{halo}^{(3)}}{M_{\sun}})$\\
           (1) & (2) &   (3)    &    (4)    &     (5)   &    (6)    &   (7)      & (8)                               &    (9)                                 & (10) & (11) & (12)\\        
\hline
470338&	n& 5& 0.09&	$25.04\pm2.56$ & $27.04\pm2.84$ & $21.17\pm2.40$ &	6.49 &	10.19 &	11.55 &	11.61 &	11.57	\\
474123&	y& 5& 0.09&	$18.92\pm2.72$ & $16.27\pm3.99$ & $14.73\pm1.75$ &	6.82 &	10.08 &	11.69 &	11.74 &	11.69	\\
474359&	y& 2& 0.30&	$23.49\pm3.71$ & $17.74\pm1.67$ & $15.34\pm2.59$ &	6.92 &	10.28 &	11.70 &	11.74 &	11.70	\\
474410&	y& 3& 0.07&	$15.89\pm1.17$ & $18.64\pm1.35$ & $21.59\pm2.66$ &	7.02 &	10.16 &	11.75 &	11.78 &	11.73	\\
475081&	n& 2& 0.33&	$16.12\pm2.27$ & $14.13\pm3.12$ & $12.95\pm3.37$ &	7.06 &	10.18 &	11.69 &	11.73 &	11.65	\\
475122&	y& 6& 0.06&	$38.00\pm1.95$ & $29.95\pm3.39$ & $35.60\pm3.88$ &	6.51 &	10.21 &	11.56 &	11.62 &	11.58	\\
475149&	n& 4& 0.10&	$15.39\pm2.94$ & $13.19\pm2.91$ & $21.40\pm5.24$ &	7.11 &	10.12 &	11.72 &	11.75 &	11.72	\\
478250&	n& 5& 0.03&	$29.62\pm3.02$ & $29.58\pm2.33$ & $28.16\pm3.41$ &	6.98 &	10.24 &	11.58 &	11.64 &	11.61	\\
478741&	n& 4& 0.18&	$24.89\pm2.65$ & $24.03\pm3.30$ & $26.45\pm3.33$ &	6.85 &	10.20 &	11.62 &	11.68 &	11.65	\\
479639&	n& 4& 0.03&	$28.21\pm3.35$ & $24.17\pm1.93$ & $23.89\pm1.88$ &	7.27 &	10.14 &	11.67 &	11.71 &	11.67	\\
480862&	n& 5& 0.14&	$19.05\pm2.45$ & $18.53\pm2.39$ & $18.42\pm1.74$ &	6.90 &	10.09 &	11.59 &	11.65 &	11.66	\\
484500&	n& 2& 0.14&	$20.03\pm2.20$ & $22.26\pm2.44$ & $19.66\pm2.40$ &	7.07 &	10.23 &	11.64 &	11.68 &	11.62	\\
484893&	n& 2& 0.10&	$19.36\pm2.88$ & $18.85\pm3.26$ & $22.50\pm2.71$ &	6.88 &	10.13 &	11.60 &	11.64 &	11.63	\\
487035&	y& 5& 0.03&	$29.60\pm3.77$ & $30.11\pm2.19$ & $22.63\pm2.63$ &	6.84 &	10.09 &	11.50 &	11.56 &	11.50	\\
487887&	n& 5& 0.16&	$30.34\pm3.82$ & $27.40\pm3.71$ & $28.92\pm3.80$ &	6.22 &	10.11 &	11.50 &	11.58 &	11.54	\\
488002&	y& 4& 0.13&	$29.43\pm3.18$ & $29.33\pm2.54$ & $29.41\pm2.72$ &	6.10 &	10.14 &	11.50 &	11.58 &	11.55	\\
488826&	y& 2& 0.09&	$17.42\pm1.53$ & $17.56\pm1.92$ & $15.35\pm2.46$ &	6.87 &	10.12 &	11.62 &	11.66 &	11.62	\\
489973&	n& 6& 0.06&	$29.76\pm3.38$ & $28.53\pm3.26$ & $29.50\pm3.92$ &	6.80 &	10.15 &	11.48 &	11.54 &	11.53	\\
491764&	n& 3& 0.04&	$23.95\pm2.85$ & $23.22\pm2.06$ & $24.11\pm2.93$ &	6.10 &	10.05 &	11.48 &	11.56 &	11.52	\\
492528&	n& 2& 0.09&	$20.13\pm2.72$ & $21.20\pm2.97$ & $30.20\pm3.95$ &	6.26 &	10.53 &	11.36 &	11.44 &	11.47	\\
493024&	n& 3& 0.22&	$26.17\pm2.65$ & $23.69\pm2.11$ & $19.35\pm2.75$ &	7.06 &	10.15 &	11.61 &	11.64 &	11.58	\\
493332&	n& 3& 0.10&	$18.64\pm1.36$ & $17.55\pm1.81$ & $18.47\pm2.57$ &	7.06 &	10.13 &	11.63 &	11.66 &	11.63	\\
494358&	n& 3& 0.07&	$13.96\pm2.19$ & $13.58\pm1.35$ & $19.72\pm3.32$ &	7.21 &	10.11 &	11.61 &	11.64 &	11.59	\\
494540&	n& 3& 0.10&	$17.63\pm1.77$ & $22.04\pm2.29$ & $24.33\pm2.25$ &	6.63 &	10.08 &	11.55 &	11.60 & 11.56	\\
494646&	n& 2& 0.16&	$26.69\pm2.80$ & $27.08\pm2.54$ & $24.65\pm2.97$ &	6.71 &	10.15 &	11.52 &	11.58 &	11.54	\\
497326&	y& 3& 0.10&	$20.29\pm3.60$ & $22.02\pm2.47$ & $23.59\pm2.39$ &	6.66 &	10.14 &	11.46 &	11.53 &	11.51	\\
498275&	n& 4& 0.03&	$23.98\pm2.60$ & $24.93\pm2.54$ & $24.20\pm2.76$ &	6.95 &	10.12 &	11.53 &	11.56 &	11.53	\\
501185&	n& 4& 0.05&	$20.07\pm2.69$ & $24.06\pm1.80$ & $26.24\pm1.84$ &	6.65 &	10.12 &	11.47 &	11.53 &	11.53	\\
503382&	y& 3& 0.10&	$28.60\pm4.40$ & $26.64\pm3.20$ & $33.69\pm2.89$ &	6.13 &	10.16 &	11.37 &	11.46 &	11.44	\\
506252&	n& 6& 0.07&	$25.98\pm3.01$ & $29.48\pm3.11$ & $31.74\pm3.22$ &	6.22 &	10.08 &	11.45 &	11.51 &	11.47	\\
508079&	n& 5& 0.03&	$28.40\pm1.99$ & $29.06\pm2.14$ & $32.23\pm2.66$ &	6.60 &	10.16 &	11.39 &	11.46 &	11.46	\\
510148&	n& 3& 0.07&	$29.32\pm2.91$ & $28.30\pm2.50$ & $27.73\pm2.51$ &	6.97 &	10.17 &	11.36 &	11.44 &	11.43	\\
518358&	n& 4& 0.02&	$23.58\pm2.65$ & $25.04\pm2.29$ & $26.94\pm2.85$ &	6.71 &	10.14 &	11.33 &	11.40 &	11.39	\\
522537&	n& 4& 0.03&	$31.95\pm2.68$ & $29.21\pm4.36$ & $28.26\pm2.45$ &	6.36 &	10.14 &	11.26 &	11.34 & 11.34	\\
\hline
\end{tabular}
\end{table*}




\bibliographystyle{mn2e}
\bibliography{reference} 


\bsp	
\label{lastpage}
\end{document}